\PassOptionsToPackage{unicode}{hyperref}
\PassOptionsToPackage{hyphens}{url}
\PassOptionsToPackage{dvipsnames,svgnames,x11names}{xcolor}
\documentclass[
  12pt]{article}

\RequirePackage{amsthm,amsmath,amsfonts,amssymb}
\RequirePackage{graphicx}
\usepackage{amsthm}
\newtheorem{theorem}{Theorem}
\numberwithin{theorem}{section}

\usepackage{graphicx} % Required for inserting images
\usepackage{url}
\usepackage{multicol}
\usepackage{xcolor}
\usepackage{calligra}
\usepackage{bm}
\usepackage{amsmath}
\usepackage{bbm}
\usepackage{dsfont}
\usepackage{algorithmicx}
\usepackage{algpseudocode}
\usepackage{hyperref}
\usepackage[ruled,vlined]{algorithm2e}
\usepackage{dsfont}
\usepackage{algorithmicx}
\usepackage{float}
\usepackage{booktabs}
\usepackage{threeparttable}

\usepackage{amsmath,amssymb}

\usepackage{iftex}
\ifPDFTeX
  \usepackage[T1]{fontenc}
  \usepackage[utf8]{inputenc}
  \usepackage{textcomp} % provide euro and other symbols
\else % if luatex or xetex
  \usepackage{unicode-math}
  \defaultfontfeatures{Scale=MatchLowercase}
  \defaultfontfeatures[\rmfamily]{Ligatures=TeX,Scale=1}
\fi
\usepackage{lmodern}
\ifPDFTeX\else  
\fi
\IfFileExists{upquote.sty}{\usepackage{upquote}}{}
\IfFileExists{microtype.sty}{% use microtype if available
  \usepackage[]{microtype}
  \UseMicrotypeSet[protrusion]{basicmath} % disable protrusion for tt fonts
}{}
\makeatletter
\@ifundefined{KOMAClassName}{% if non-KOMA class
  \IfFileExists{parskip.sty}{%
    \usepackage{parskip}
  }{% else
    \setlength{\parindent}{0pt}
    \setlength{\parskip}{6pt plus 2pt minus 1pt}}
}{% if KOMA class
  \KOMAoptions{parskip=half}}
\makeatother
\usepackage{xcolor}
\makeatletter
\ifx\paragraph\undefined\else
  \let\oldparagraph\paragraph
  \renewcommand{\paragraph}{
    \@ifstar
      \xxxParagraphStar
      \xxxParagraphNoStar
  }
  \newcommand{\xxxParagraphStar}[1]{\oldparagraph*{#1}\mbox{}}
  \newcommand{\xxxParagraphNoStar}[1]{\oldparagraph{#1}\mbox{}}
\fi
\ifx\subparagraph\undefined\else
  \let\oldsubparagraph\subparagraph
  \renewcommand{\subparagraph}{
    \@ifstar
      \xxxSubParagraphStar
      \xxxSubParagraphNoStar
  }
  \newcommand{\xxxSubParagraphStar}[1]{\oldsubparagraph*{#1}\mbox{}}
  \newcommand{\xxxSubParagraphNoStar}[1]{\oldsubparagraph{#1}\mbox{}}
\fi
\makeatother

\usepackage{longtable,booktabs,array}
\usepackage{calc} % for calculating minipage widths
\usepackage{etoolbox}
\makeatletter
\patchcmd\longtable{\par}{\if@noskipsec\mbox{}\fi\par}{}{}
\makeatother
\IfFileExists{footnotehyper.sty}{\usepackage{footnotehyper}}{\usepackage{footnote}}
\makesavenoteenv{longtable}
\usepackage{graphicx}
\makeatletter
\def\maxwidth{\ifdim\Gin@nat@width>\linewidth\linewidth\else\Gin@nat@width\fi}
\def\maxheight{\ifdim\Gin@nat@height>\textheight\textheight\else\Gin@nat@height\fi}
\makeatother
\setkeys{Gin}{width=\maxwidth,height=\maxheight,keepaspectratio}
\makeatletter
\def\fps@figure{htbp}
\makeatother

\makeatletter
\@ifpackageloaded{caption}{}{\usepackage{caption}}
\AtBeginDocument{%
\ifdefined\contentsname
  \renewcommand*\contentsname{Table of contents}
\else
  \newcommand\contentsname{Table of contents}
\fi
\ifdefined\listfigurename
  \renewcommand*\listfigurename{List of Figures}
\else
  \newcommand\listfigurename{List of Figures}
\fi
\ifdefined\listtablename
  \renewcommand*\listtablename{List of Tables}
\else
  \newcommand\listtablename{List of Tables}
\fi
\ifdefined\figurename
  \renewcommand*\figurename{Figure}
\else
  \newcommand\figurename{Figure}
\fi
\ifdefined\tablename
  \renewcommand*\tablename{Table}
\else
  \newcommand\tablename{Table}
\fi
}
\@ifpackageloaded{float}{}{\usepackage{float}}
\floatstyle{ruled}
\@ifundefined{c@chapter}{\newfloat{codelisting}{h}{lop}}{\newfloat{codelisting}{h}{lop}[chapter]}
\floatname{codelisting}{Listing}

\makeatother
\makeatletter
\@ifpackageloaded{caption}{}{\usepackage{caption}}
\@ifpackageloaded{subcaption}{}{\usepackage{subcaption}}
\makeatother

\ifLuaTeX
  \usepackage{selnolig}  % disable illegal ligatures
\fi
\usepackage[]{natbib}
\usepackage{bookmark}

\IfFileExists{xurl.sty}{\usepackage{xurl}}{} % add URL line breaks if available
\hypersetup{
  pdftitle={Title},
  pdfauthor={Author 1; Author 2},
  pdfkeywords={3 to 6 keywords, that do not appear in the title},
  colorlinks=true,
  linkcolor={blue},
  filecolor={Maroon},
  citecolor={Blue},
  urlcolor={Blue},
  pdfcreator={LaTeX via pandoc}}

\newcommand{\anon}{1}

\usepackage{amsthm}

\newtheorem{assumption}{Assumption}

\begin{document}

\def\spacingset#1{\renewcommand{\baselinestretch}%
{#1}\small\normalsize} \spacingset{1}

%%%%%%%%%%%%%%%%%%%%%%%%%%%%%%%%%%%%%%%%%%%%%%%%%%%%%%%%%%%%%%%%%%%%%%%%%%%%%%

\if1\anon
{
  \title{\bf Bayesian Transfer Learning for High-Dimensional Linear Regression via Adaptive Shrinkage}

  \author{
    Parsa Jamshidian and Donatello Telesca\\
    Department of Biostatistics, University of California, Los Angeles\\
}

\date{}

\maketitle

  \maketitle
} \fi

\bigskip
\begin{abstract}
We introduce BLAST, Bayesian Linear regression with Adaptive Shrinkage for Transfer, a Bayesian multi-source transfer learning framework for high-dimensional linear regression. The proposed analytical framework leverages global-local shrinkage priors together with Bayesian source selection to balance information sharing and regularization. We show how Bayesian source selection allows for the extraction of the most useful data sources, while discounting biasing information that may lead to negative transfer. In this framework, both source selection and sparse regression are jointly accounted for in prediction and inference via Bayesian model averaging.  The structure of our model admits efficient posterior simulation via a Metropolis-within-Gibbs sampling algorithm allowing full posterior inference for the target regression coefficients, making BLAST both computationally practical and inferentially straightforward. Our method achieves more accurate posterior inference for the target than regularization approaches based on target data alone, while offering competitive predictive performance and superior uncertainty quantification compared to current state-of-the-art transfer learning methods. We validate its effectiveness through extensive simulation studies and illustrate its analytical properties when applied to a case study on the estimation of tumor mutational burden from gene expression, using data from The Cancer Genome Atlas (TCGA).
\end{abstract}

\noindent%
\vfill

\newpage
\spacingset{1.8} % DON'T change the spacing!

\section{Introduction}

In biomedical applications such as rare disease studies and personalized medicine, sample sizes are often inherently limited, making reliable statistical inference challenging. Transfer learning (TL) encompasses a broad class of analytical approaches that aim to leverage information from one or more related domains to improve inference in a target domain \citep{Pan2010Survey, Suder2025BayesianTransferLearning}. This work investigates the multi-source transfer learning problem in the setting of high-dimensional linear regression, where a target population or application is interrogated through a target data-set $\mathcal{D}_0 = (\mathbf{X}^{(0)}, \mathbf{y}^{(0)})$ with design matrix $\mathbf{X}^{(0)} \in \mathbb{R}^{n_0 \times p}$ and outcome vector $\mathbf{y}^{(0)} \in \mathbb{R}^{n_0}$. Potentially related studies are encoded in multiple auxiliary data sources denoted as $\mathcal{D}_k = (\mathbf{X}^{(k)}, \mathbf{y}^{(k)})$, where  $\mathbf{X}^{(k)} \in \mathbb{R}^{n_k \times p}$ and $\mathbf{y}^{(k)}\in \mathbb{R}^{n_k}$ for $k = 1, \ldots, K$. 

Our overarching objective is to integrate information from multiple source datasets in order to enhance inference and predictive performance for the target task. Achieving this objective requires addressing two central statistical challenges: (1) developing a principled formalism for borrowing information across studies, and (2) identifying and selecting data sources that contribute useful signal without inducing substantial bias, thereby avoiding negative transfer.

%Transfer learning in high-dimensional linear regression has attracted increasing attention in recent years, particularly within the frequentist paradigm. 
The proposed formalism for modeling study relatedness builds on the Trans-Lasso method of  \citet{Li2022TransferOptimality}, who conceptualized TL in high-dimensional linear regression through the idea of sparse contrasts. Precisely, let $\mathcal{A}\subseteq \{1,2,\ldots,K\}$ index a subset of informative auxiliary studies. Trans-Lasso constructs the regression coefficients in the target task, say $\boldsymbol{\beta}\in \mathbb{R}^p$, as the sum of two sparse vectors: a coefficient vector obtained by pooling informative sources, say ${\bm w}\in\mathbb{R}^p$, and a vector of sparse contrasts $\boldsymbol{\delta}\in \mathbb{R}^p$, so that $\boldsymbol{\beta} = {\bm w} + \boldsymbol{\delta}$. Their estimation strategy relies on a two-stage Lasso-based estimator, and final aggregation over multiple candidate sets $\mathcal{A}$ via Q-aggregation \citep{Dai2012}. \cite{Li2022TransferOptimality} showed how the procedure achieves enhanced precision in estimation as the number of informative studies increases, but did not develop a theory for uncertainty quantification. \citet{Tian2023TransferModels} later extended this two-stage approach to high-dimensional generalized linear models and developed a selective inference procedure for constructing asymptotic confidence intervals for the target estimates.

Similar ideas have been exploited from a Bayesian perspective.   \citet{Abba2024ALearning} used the horseshoe prior \citep{Carvalho2009HandlingHorseshoe} in a single-source setting to model the contrast between source and target means in the normal means problem. For multi-source integration, \citet{Lai2024BayesianInference} proposed centering the horseshoe prior at a weighted average of pre-estimated source coefficients. Finally, \citet{Zhang2024Covariate-ElaboratedPrior} introduced extensions to multi-source TL in high-dimensional linear regression through conditional spike-and-slab priors to enable selective borrowing across sources through the use of latent covariate inclusion indicators. \citet{Suder2025BayesianTransferLearning} give a comprehensive overview of recent Bayesian approaches for TL, highlighting the role of hierarchical modeling in leveraging source information.

%\subsection{Our Contribution}

These seminal contributions have made meaningful strides in the construction of TL estimators in multi-source high-dimensional regression. However, important methodological limitations are still at play when inference is to be made on target regression coefficients after TL. Particularly, the selective inference approach of \cite{Tian2023TransferModels} relies on fixing an empirically determined informative set $\mathcal{A}$. While asymptotically valid, this procedure is likely too optimistic in most finite-sample settings --- the very situations which would warrant reliance on TL techniques (see Section 4). In the Bayesian setting, the approach of \cite{Zhang2024Covariate-ElaboratedPrior} provides a conceptually flexible and robust framework for TL. However, reliance on spike-and-slab variable and contrasts selection results in a highly complex discrete model search problem which is solved via variational approximations, and therefore provides posterior approximations which are likely not suited for uncertainty quantification.

Our proposal builds on these contributions  
and addresses the important problem of inference after TL, via a simple application of Bayesian model averaging. We name this method BLAST, which stands for \textbf{B}ayesian \textbf{L}inear regression with \textbf{A}daptive \textbf{S}hrinkage for \textbf{T}ransfer. BLAST performs \emph{study-level} borrowing and adaptively learns sparsity from individual-level data through continuous shrinkage. Sparse estimation is compatible with a broad class of global-local shrinkage priors \citep{Bahdra2016}, providing flexibility to tailor prior specifications to different problem contexts or domain knowledge. When the informative source set $\mathcal{A}$ is unknown, we introduce latent source study--level inclusion indicators and infer $\mathcal{A}$ jointly with model parameters through posterior sampling, with inference carried out via Bayesian model averaging over probable configurations of $\mathcal{A}$. While eminently Bayesian, this procedure is shown to have good theoretical and empirical frequentist properties.

The rest of our paper is organized as follows. In Section~2, we review the general Bayesian shrinkage model and present the methodology of BLAST in detail for the $\mathcal{A}$-known and $\mathcal{A}$-unknown cases. We further provide an example of the model specification and implementation using the horseshoe shrinkage prior. In Section 3, we present large-sample theoretical guarantees for our method. We show results of simulations using our method in Section 4, comparing the performance with other popular transfer learning algorithms in the literature. In Section 5, we demonstrate the efficacy of our method in a real-data application which involves the prediction of tumor mutational burden from gene expression profiles using publicly available data from The Cancer Genome Atlas (TCGA). Finally, in Section 6 we summarize our findings and provide concluding remarks. 

\section{Methodology}

In this section, 
we introduce the multi-source transfer learning framework, BLAST, under both the oracle setting, where the informative set $\mathcal{A}$ is taken to be known, and the more practical setting in which $\mathcal{A}$ must be inferred from the data. In each case, 
we discuss posterior inference via Monte Carlo sampling under general global-local shrinkage priors.
An example implementation of the BLAST framework using the popular horseshoe shrinkage prior of \cite{Carvalho2010TheSignals} is finally introduced for clarity and reproducibility.

\subsection{Oracle BLAST: The $\mathcal{A}$-known Case}

We first describe BLAST under the assumption that the informative set $\mathcal{A}$ is assumed to be known in advance. This setting is referred to as the \emph{oracle} case
and serves as an important ideal benchmark. 

Consider a target dataset $\mathcal{D}_0 = \bigl(\mathbf{y}^{(0)}, \mathbf{X}^{(0)}\bigr)$, where $\mathbf{y}^{(0)} \in \mathbb{R}^{n_0}$ is the outcome vector and $\mathbf{X}^{(0)} \in \mathbb{R}^{n_0 \times p}$ is the corresponding design matrix.  
In addition, we have a collection of $K$ informative source datasets indexed by $\mathcal{A} = \{1,2,\ldots,K\}$, where each dataset is given by $\mathcal{D}_k = \bigl(\mathbf{y}^{(k)}, \mathbf{X}^{(k)}\bigr)$ with $\mathbf{y}^{(k)} \in \mathbb{R}^{n_k}$ and $\mathbf{X}^{(k)} \in \mathbb{R}^{n_k \times p}$ for $k = 1,\ldots,K$. Further, let
\[
\mathbf{X}^{(\mathcal{A})} = \begin{bmatrix}
    \mathbf{X}^{(1)} \\ 
    \mathbf{X}^{(2)} \\ 
    \vdots \\
    \mathbf{X}^{(K)}
\end{bmatrix}, \qquad
\mathbf{y}^{(\mathcal{A})} = \begin{bmatrix}
\mathbf{y}^{(1)} \\
\mathbf{y}^{(2)} \\
\vdots \\
\mathbf{y}^{(K)}
\end{bmatrix},
\]
be the stacked design matrix and outcome vector of the informative source datasets, respectively. Here, $\mathbf{X}^{(\mathcal{A})} \in \mathbb{R}^{n_\mathcal{A} \times p}$ and $\mathbf{y}^{(\mathcal{A})} \in \mathbb{R}^{n_\mathcal{A}}$, where $n_\mathcal{A} = \sum_{k=1}^K n_k$.

The sampling model assumes:
\begin{equation} \label{Aknownlik} 
\begin{array}{lclcl}
\mathbf{y}^{(\mathcal{A})} &\mid& \mathbf{X}^{(\mathcal{A})},\; \bm{w}^{(\mathcal{A})},\; \sigma^2_{(\mathcal{A})} &\sim& \mathcal{N}\left(\mathbf{X}^{(\mathcal{A})} \bm{w}^{(\mathcal{A})},\; \sigma^2_{(\mathcal{A})} \mathbf{I}\right), \\
\mathbf{y}^{(0)} &\mid& \mathbf{X}^{(0)},\; \bm{w}^{(\mathcal{A})},\; \boldsymbol{\delta},\; \sigma^2_{(0)} &\sim& \mathcal{N}\left\{\mathbf{X}^{(0)} \left( \bm{w}^{(\mathcal{A})} + \boldsymbol{\delta} \right),\; \sigma^2_{(0)} \mathbf{I} \right\},
\end{array}
\end{equation}
where we require that all auxiliary data sets are anchored to the target through the coefficients ${\bm w}^{(\mathcal{A})}$, and the target is allowed to deviate from ${\bm w}^{(\mathcal{A})}$ via a set of sparse contrasts $\boldsymbol{\delta}$. In other words, the target regression coefficient takes the form $\boldsymbol{\beta} = \bm{w}^{(\mathcal{A})} + \boldsymbol{\delta}$ where $\bm{w}^{(\mathcal{A})}$ and $\boldsymbol{\delta}$ are aggregate regression parameters representing the source data coefficients and contrasts, respectively.

Under sparsity for both regression coefficients and contrasts, a natural prior model may then rely on independent Normal scale-mixture  priors of the form given in \citep{Bahdra2016}, s.t.:
\begin{equation}
{\bm w}^{(\mathcal{A})}_j \mid \sigma^2_{(\mathcal{A})}, \nu_j^{\bm w}
\sim \mathcal{N}\!\left(0,\;\sigma^2_{(\mathcal{A})}\nu_j^{\bm w}\right),\qquad
\boldsymbol{\delta}_j \mid \sigma^2_{(0)}, \nu_j^{\bm \delta}
\sim \mathcal{N}\!\left(0,\;\sigma^2_{(0)}\nu_j^{\bm \delta}\right).
\label{eq:shrinkage_prior_normal}
\end{equation}

Here, each regression coefficient is assigned its own set of local shrinkage parameters, $\bm\nu^{\bm w} = \{\nu_j^{\bm w}\}_{j = 1}^p$ and $\bm\nu^{\bm \delta} = \{\nu_j^{\bm \delta}\}_{j = 1}^p$, whose prior distribution, in turn, defines the shrinkage topology after marginalization. For example, under the horseshoe prior \citep{Carvalho2010TheSignals}, one may specify
\[
\nu_j^{\bm w} = \lambda_j^2 \tau^2,
\qquad
\lambda_j \sim \mathrm{C}^+(0,1),
\qquad
\tau \sim \mathrm{C}^+(0,1),
\]
with the global shrinkage parameter $\tau^2$ controlling the number of signals, and local shrinkage parameters $\lambda_j^2$ selecting the signal coefficients allowed by $\tau^2$. The contrasts prior for 
$\nu_j^{\bm \delta}$ is defined analogously for $j = 1, \ldots, p$. A review of alternative global-local shrinkage parametrizations is provided in supplemental Appendix A. 
% let  $\bm\nu^{\bm w} = \{\nu_j^{\bm w}\}_{j = 1}^p$ and $\bm\nu^{\bm \delta} = \{\nu_j^{\bm \delta}\}_{j = 1}^p$ denote the corresponding vector of coefficients and contrasts prior variances. 

Finally, we place inverse-gamma priors on the residual variances,
\[
\sigma^2_{(\mathcal{A})},\; \sigma^2_{(0)} \sim \mathrm{IG}(a,b), \qquad a,b>0,
\]
reflecting a weakly-informative, heavy-tailed prior on the residual scale. 

Taken together, the likelihood in \eqref{Aknownlik} and the shrinkage priors in \eqref{eq:shrinkage_prior_normal} define a fully specified hierarchical Bayesian model. Let $\bm\vartheta = \left(\bm w^{(\mathcal{A})}, \boldsymbol{\delta}, 
\bm \nu^{\bm w}, \bm \nu^{\bm \delta}, 
\sigma^2_{(\mathcal{A})}, \sigma^2_{(0)}\right)$ 
denote the full collection of unknown parameters and let 
$\mathcal{D} = \{\mathcal{D}_0, \mathcal{D}_{\mathcal{A}}\}$ denote the observed data from the target and informative source studies. 
By Bayes' theorem, the posterior density satisfies
\begin{equation}
p(\bm\vartheta \mid \mathcal{D})
\;\propto\;
L_{\mathcal{A}}\!\left(\mathcal{D}_{\mathcal{A}}
\mid \bm w^{(\mathcal{A})}, \sigma^2_{(\mathcal{A})}\right)
\,
L_{0}\!\left(\mathcal{D}_{0}
\mid \bm w^{(\mathcal{A})}, \boldsymbol{\delta}, \sigma^2_{(0)}\right)
\,
\pi(\bm\vartheta).
\label{eq:posterior_compact_Aknown}
\end{equation}

where $L_{\mathcal{A}}$ and $L_0$ denote the Gaussian likelihood contributions
from the informative sources and the target data, respectively, and
$\pi(\bm\vartheta)$ denotes the joint prior distribution induced by the
continuous shrinkage hierarchy on the regression coefficients together with
the hyperpriors on the associated shrinkage parameters and error variances.
Posterior inference for the target coefficients 
$\boldsymbol{\beta} = \bm w^{(\mathcal{A})} + \boldsymbol{\delta}$ 
can be carried out via Markov Chain Monte Carlo (MCMC) sampling from the posterior distribution in \eqref{eq:posterior_compact_Aknown}.

\subsection{Oracle BLAST Algorithm}
The hierarchical model in (\ref{eq:posterior_compact_Aknown}) admits a convenient Metropolis-within-Gibbs sampling procedure for obtaining joint posterior samples of the parameters of interest. Algorithm~\ref{algAknown} outlines the proposed sampler for estimating the target regression coefficients $\bm{\beta}$ in the oracle version of the BLAST framework, which we refer to as Oracle BLAST. The algorithm jointly samples the source coefficients $\bm{w}^{(\mathcal{A})}$ and the contrast vector $\bm{\delta}$ from their respective full conditional distributions, with posterior samples of $\bm{\beta}$ obtained as the sum $\bm{w}^{(\mathcal{A})} + \bm{\delta}$.  

The full conditional distributions for both the anchoring coefficients and the contrast parameters are Gaussian, yielding conjugate updates within the Gibbs framework. For example, the conditional distribution of the shared anchoring coefficients, $\bm{w}^{(\mathcal{A})}$, takes the form
\[
\bm{w}^{(\mathcal{A})} \mid \text{rest} \sim \mathcal N(\bm\mu_w, \bm\Lambda_w^{-1}),
\]
where the precision matrix and mean vector are given by
\begin{align*}
\bm\Lambda_w
&=
\frac{1}{\sigma^2_{(\mathcal A)}}\!\left(\mathbf X^{(\mathcal A)\top}\mathbf X^{(\mathcal A)} + \mathbf D_{(\mathcal A)}^{-1}\right)
\;+\;
\frac{1}{\sigma^2_{(0)}}\,\mathbf X^{(0)\top}\mathbf X^{(0)},
\\[6pt]
\bm\mu_w
&=
\bm\Lambda_w^{-1}
\!\left\{
\frac{1}{\sigma^2_{(\mathcal A)}}\,\mathbf X^{(\mathcal A)\top}\mathbf y^{(\mathcal A)}
\;+\;
\frac{1}{\sigma^2_{(0)}}\,\mathbf X^{(0)\top}\!\left(\mathbf y^{(0)} - \mathbf X^{(0)}\bm\delta\right)
\right\},
\end{align*}
and $\mathbf{D}_{(\mathcal{A})}$ is a diagonal matrix containing the local shrinkage parameters associated with the source coefficients. The full conditional distribution for the contrast vector, $\bm\delta \mid \text{rest}$, admits an analogous Gaussian form conditional on the anchoring coefficients, and conjugate updates are also available for the residual variance parameters (see Appendix~B). In high-dimensional regression, this model can leverage efficient sampling for the regression parameters such as that in \citet{Bhattacharya2016FastRegression} for Gaussian scale-mixture priors, which has computational complexity $\mathcal{O}(n^2p)$ and is well-suited for large $p$ settings. 

The shrinkage parameters themselves need not admit closed-form full conditionals. Their updates depend on the chosen hyperpriors and may be implemented using Gibbs, Metropolis--Hastings (MH), or accept--reject (AR) steps within the overall Gibbs scheme. One may specify $\nu_j^{\bm w}$ and $\nu_j^{\bm\delta}$ to share the same prior form allowing a common sampling routine or assign distinct priors to reflect differing beliefs about sparsity in the anchoring and contrast components.

\begin{algorithm}[t]
\footnotesize
\caption{Metropolis-within-Gibbs sampler for Oracle BLAST}\label{algAknown}
\KwIn{Target Data $\mathcal{D}_0$, Informative Source Data $\mathcal{D}_{\mathcal{A}}$, No.\ of MCMC iterations $T$}
\KwOut{Posterior samples $\{\bm\beta^{(t)}, \bm\nu_{\bm w}^{(t)}, \bm\nu_{\bm\delta}^{(t)}, \sigma_{(0)}^{2(t)}, \sigma_{(\mathcal{A})}^{2(t)}\}_{t=1}^{T}$}

\textbf{Initialize:} $\bm{w}^{(\mathcal{A})}, \bm\delta \gets \mathbf{0}_p$; $\sigma^2_{(0)}, \sigma^2_{(\mathcal{A})} \gets 1$; $\bm\nu_{\bm w}, \bm\nu_{\bm\delta} \gets \mathbf{1}_p$\;

\For{$t=1$ \KwTo $T$}{
  \[
  \begin{array}{rclcl}
  \bm{w}^{(\mathcal{A})}
  &\mid&
  \mathcal{D}_0, \mathcal{D}_{\mathcal{A}}, \bm\delta, \bm\nu_{\bm w}, \sigma^2_{(\mathcal{A})}
  &\sim&
  \mathcal{N}\!\big(\bm\mu_w,\; \mathbf{\Lambda}_w^{-1}\big)
  \\[6pt]
  \bm\delta
  &\mid&
  \mathcal{D}_0, \bm{w}^{(\mathcal{A})}, \bm\nu_{\bm\delta}, \sigma^2_{(0)}
  &\sim&
  \mathcal{N}\!\big(\bm\mu_\delta,\; \mathbf{\Lambda}_\delta^{-1}\big)
  \\[6pt]
  \sigma^{2}_{(0)}
  &\mid&
  \mathcal{D}_0, \bm{w}^{(\mathcal{A})}, \bm\delta
  &\sim&
  \mathrm{InvGamma}\!\big(a_0^*,\; b_0^*\big)
  \\[6pt]
  \sigma^{2}_{(\mathcal{A})}
  &\mid&
  \mathcal{D}_{\mathcal{A}}, \bm{w}^{(\mathcal{A})}
  &\sim&
  \mathrm{InvGamma}\!\big(a_{\mathcal{A}}^*,\; b_{\mathcal{A}}^*\big)
  \end{array}
  \]

  Update shrinkage parameters $\bm\nu_{\bm w}, \bm\nu_{\bm\delta}$ via Gibbs, MH or AR step\;

  Compute $\bm\beta = \bm{w}^{(\mathcal{A})} + \bm\delta$\;

  Store $(\bm\beta, \bm\nu_{\bm w}, \bm\nu_{\bm\delta}, \sigma^2_{(0)}, \sigma^2_{(\mathcal{A})})$\;
}
\KwRet{$\{\bm\beta^{(t)}, \bm\nu_{\bm w}^{(t)}, \bm\nu_{\bm\delta}^{(t)}, \sigma_{(0)}^{2(t)}, \sigma_{(\mathcal{A})}^{2(t)}\}_{t=1}^{T}$}\;
\end{algorithm}

\subsection{BLAST with Source Selection: The $\mathcal{A}$-unknown Case}

In many practical applications, the informative set $\mathcal{A}$ is not known a priori. Importantly, failing to correctly identify $\mathcal{A}$ and naively incorporating noninformative source studies can lead to negative transfer and degraded performance \citep{Li2022TransferOptimality}. To address this, we extend the $\mathcal{A}$-known model by introducing a latent $K$-dimensional binary indicator vector $\bm\gamma = (\gamma_1, \ldots, \gamma_K)$, where each $\gamma_k \in \{0,1\}$ represents whether the $k$-th source contributes transferable information to the target task.

Rather than fixing $\mathcal{A}$, we infer $\bm\gamma$ jointly with the model parameters, allowing the degree of information sharing to be learned adaptively from the data. Posterior inference is carried out using a Metropolis-within-Gibbs scheme that integrates source selection with shrinkage-based estimation.

In this more general case, we consider a library $\mathcal{S} := \{1,2,\ldots,K\}$ of $K$ available source data sets from which we aim to identify the informative set $\mathcal{A} \subseteq \mathcal{S}$. We introduce a $K$-dimensional latent binary indicator vector $\bm\gamma = (\gamma_1, \ldots, \gamma_K)\sim f_\gamma$, with $f_\gamma$ denoting a probability mass function supported on $\Gamma := \{0,1\}^K$ and factorizing as
\begin{equation}
f_{\bm\gamma}(\bm\gamma \mid \pi)
= \prod_{k=1}^K \pi^{\gamma_k}(1-\pi)^{1-\gamma_k},
\qquad \bm\gamma \in \Gamma,
\label{eq:fgamma_bern}
\end{equation}
corresponding to $\gamma_k \mid \pi \stackrel{\mathrm{ind}}{\sim} \mathrm{Bernoulli}(\pi)$ for $k=1,\ldots,K$. In this framework, we assume that any realization of $\bm\gamma\in \Gamma$ partitions the $K$ source datasets into:
\begin{center}
\begin{tabular}{llcl}
{\bf Informative Sources} & ${\mathcal{A}}_{\bm\gamma}$ &=& $\{k\in(1,2,\ldots,K): \gamma_k = 1\}$,\\[0.05in]
{\bf Noninformative Sources} & ${\bar{\mathcal{A}}_{\bm\gamma}}$ &=& $\{k\in(1,2,\ldots,K): \gamma_k = 0\}.$
\end{tabular}
\end{center}
%Henceforth, we denote noninformative data as $\DgammaAbar = (\mathbf{X}^{(\bar{\mathcal{A}}_{\bm\gamma})},\mathbf{y}^{(\bar{\mathcal{A}}_{\bm\gamma})})$.
Informative sources are expected to contribute positively to inference as they are deemed compatible with a sparse contrast structure, whereas noninformative sources may degrade performance. 

To account for uncertainty in the informative source set, we extend the transfer learning model in \eqref{Aknownlik} by introducing dependence on the latent source membership vector $\bm{\gamma}$. 
Conditional on $\bm{\gamma}$, the sampling model in the $\mathcal{A}$-unknown case is specified as
\begin{equation}
\label{eq:Aunknownlik}
\begin{aligned}
\mathbf{y}^{(\mathcal{A}_{\bm\gamma})} 
&\mid \mathbf{X}^{(\mathcal{A}_{\bm\gamma})},\, \bm{w}^{(\mathcal{A})},\, \sigma^2_{(\mathcal{A})}
\sim \mathcal{N}\!\left(
\mathbf{X}^{(\mathcal{A}_{\bm\gamma})} \bm{w}^{(\mathcal{A})},\,
\sigma^2_{(\mathcal{A})}\mathbf{I}
\right), \\
\mathbf{y}^{(0)} 
&\mid \mathbf{X}^{(0)},\, \bm{w}^{(\mathcal{A})},\, \boldsymbol{\delta},\, \sigma^2_{(0)}
\sim \mathcal{N}\!\left\{
\mathbf{X}^{(0)}(\bm{w}^{(\mathcal{A})} + \boldsymbol{\delta}),\,
\sigma^2_{(0)}\mathbf{I}
\right\}, \\
\mathbf{y}^{(\bar{\mathcal{A}}_{\bm\gamma})} 
&\mid \mathbf{X}^{(\bar{\mathcal{A}}_{\bm\gamma})},\, \bm{w}^{(\bar{\mathcal{A}})},\, \sigma^2_{(\bar{\mathcal{A}})}
\sim \mathcal{N}\!\left(
\mathbf{X}^{(\bar{\mathcal{A}}_{\bm\gamma})} \bm{w}^{(\bar{\mathcal{A}})},\,
\sigma^2_{(\bar{\mathcal{A}})}\mathbf{I}
\right),
\end{aligned}
\end{equation}

where we have introduced an additional normal likelihood component for noninformative sources. 
We retain the shrinkage priors specified in \eqref{eq:shrinkage_prior_normal} for the informative sources and contrast parameters, and additionally place a shrinkage prior of the same form on the noninformative source coefficients:
\[
w^{(\bar{\mathcal{A}})}_j 
\mid \sigma^2_{(\bar{\mathcal{A}})}, \nu_j^{\bar{\bm w}}
\sim
\mathcal{N}\!\left(
0,\,
\sigma^2_{(\bar{\mathcal{A}})} \nu_j^{\bar{\bm w}}
\right).
\]
Although inference on $\bm{w}^{(\bar{\mathcal{A}})}$ is not of primary interest, its shrinkage parameters play a critical role in evaluating source compatibility and learning the latent vector $\bm{\gamma}$ as we will demonstrate later. Additional stochastic constraints may be placed on the contrast vector $\boldsymbol{\delta}$ to ensure tight compatibility with the informative set $\mathcal{A}_{\boldsymbol{\gamma}}$. Specifically, we may ask that the sparsity in $\boldsymbol{\delta}$ exceeds the sparsity in ${\bm w}^{(\mathcal{A})}$. We discuss the details of these constraints in Section \ref{HorseshoeImplementation}.

The posterior distribution of the target parameter $\boldsymbol{\beta}$ is naturally represented as
$$p(\boldsymbol{\beta}\mid \mathcal{D}) = \sum_{\boldsymbol{\gamma}\in \Gamma} p(\boldsymbol{\beta}\mid \mathcal{D}, \boldsymbol{\gamma})p(\boldsymbol{\gamma}\mid \mathcal{D}).$$
Here, inferential uncertainty in the selection of the informative set $\mathcal{A}$ is made explicit through Bayesian model averaging over $\Gamma$. Crucially, the posterior distribution $p(\boldsymbol{\gamma}\mid \mathcal{D})$ induces a posterior over all candidate informative sets $\mathcal{A}\subseteq\mathcal{S}$, and thus posterior uncertainty regarding which sources are informative is fully characterized by $p(\boldsymbol{\gamma}\mid \mathcal{D})$, which assigns posterior probabilities  to each of the $2^K$ candidate configurations in $\Gamma$.

\subsection{BLAST Algorithm with Source Selection}
When the informative set $\mathcal{A}$ is unknown, posterior inference must be carried out jointly over both the regression parameters and the latent source inclusion indicators. This introduces an additional layer of uncertainty relative to the oracle setting, as the model must explore different candidate subsets of informative studies during sampling. Consequently, we extend the Metropolis-within-Gibbs sampler described in Algorithm~\ref{algAknown} to incorporate updates of the binary inclusion vector $\boldsymbol{\gamma}$.

Algorithm~\ref{algAunknown} presents our Metropolis-within-Gibbs sampling procedure for obtaining posterior inference on the target regression coefficients when \(\mathcal{A}\) is unknown. Although most of the underlying sampling machinery is similar, there are several key differences to note between Algorithm~\ref{algAunknown} and Algorithm~\ref{algAknown}. 

The first distinction arises from the introduction of an additional likelihood component corresponding to the noninformative sources. Algorithm~\ref{algAunknown} includes updates for the regression
parameters associated with noninformative sources, 
$\bm w^{(\bar{\mathcal{A}})}$. These updates retain the same conjugate normal structure as those for
$\bm w^{(\mathcal{A})}$ and add minimal computational complexity.
More fundamentally, because the informative set is no longer fixed, the partition of studies depends on the current configuration $\boldsymbol{\gamma}$ through the induced sets $\mathcal{A}_{\boldsymbol{\gamma}}$ and $\bar{\mathcal{A}}_{\boldsymbol{\gamma}}$. As a result, the posterior conditional means and precision matrices for the regression parameters must be re-evaluated at each iteration using the data assigned to these sets.

The second critical difference is that each iteration of the 
Metropolis-within-Gibbs sampler in Algorithm~\ref{algAunknown} includes a 
\emph{source study selection step}, which updates the latent source inclusion vector 
$\boldsymbol{\gamma}$. Specifically, the algorithm includes a single-site Metropolis--Hastings update by 
sequentially proposing flips $\gamma_k' = 1-\gamma_k$ for $k = 1,\ldots,K$. 
Each proposal is accepted with probability determined by the ratio of posterior 
densities under the proposed and current configurations. Evaluating this update requires computing
$p(\boldsymbol{\gamma}\mid\boldsymbol{\nu},\mathcal{D})$, which depends on the
marginal likelihood under a fixed configuration of $\boldsymbol{\gamma}$. An explicit expression for this quantity is derived in Appendix~B. The same Appendix section also discusses a burn-in tempering strategy and numerical stability details.

%Obtaining this marginal likelihood involves integrating out the regression parameters with respect to their shrinkage priors, yielding a high-dimensional integral that must be handled carefully to maintain computational tractability. We therefore briefly detail this below.

%\paragraph*{Marginal Likelihood for Source Study Selection Step}

\begin{algorithm}[t!]
\footnotesize
\caption{Metropolis-within-Gibbs sampler for BLAST with Source Selection}\label{algAunknown}
\KwIn{Target Data $\{\mathbf{X}^{(0)}, \mathbf{y}^{(0)}\}$, Source Data $\{\mathbf{X}^{\mathcal{(S)}}, \mathbf{y}^{\mathcal{(S)}}\}$, No. of MCMC iterations $T$}
\KwOut{Posterior samples $\{\bm\beta^{(t)}, \bm\gamma^{(t)}, \bm\nu^{(t)}, \bm\sigma^{2(t)}\}_{t=1}^{T}$, Inclusion Probabilities $\{\sum_{t=1}^T \gamma_k^{(t)} / T\}_{k=1}^K$}

\textbf{Initialize:} 
$\bm{w}^{(\mathcal{A})}, \bm{w}^{(\bar{\mathcal{A}})}, \bm\delta \gets \mathbf{0}_p$; 
$\bm\gamma \gets \mathbf{1}_K$; 
$\sigma^2_{(0)}, \sigma^2_{(\mathcal{A})}, \sigma^2_{(\bar{\mathcal{A}})} \gets 1$; 
$\bm\nu_{\bm{w}}, \bm\nu_{\bar{\bm{w}}}, \bm\nu_{\bm\delta} \gets \mathbf{1}_d$.

\For{$t = 1$ \KwTo $T$}{
  
\[
\begin{array}{rclcl}
\bm{w}^{(\mathcal{A})}
&\mid&
\mathcal{D}_0,\; \mathcal{D}_{\mathcal{A}_{\boldsymbol{\gamma}}},\;
\bm\delta,\; \bm\nu_{\bm{w}},\; \sigma^2_{(\mathcal{A})},\;
\boldsymbol{\gamma}
&\sim&
\mathcal{N}\!\big(\bm\mu_{w}^{(\boldsymbol{\gamma})},\;
(\mathbf{\Lambda}_{w}^{(\boldsymbol{\gamma})})^{-1}\big)
\\[6pt]
\bm\delta
&\mid&
\mathcal{D}_0,\; \bm{w}^{(\mathcal{A})},\;
\bm\nu_{\bm\delta},\; \sigma^2_{(0)},\;
\boldsymbol{\gamma}
&\sim&
\mathcal{N}\!\big(\bm\mu_{\delta},\;
\mathbf{\Lambda}_{\delta}^{-1}\big)
\\[6pt]
\bm{w}^{(\bar{\mathcal{A}})}
&\mid&
\mathcal{D}_{\bar{\mathcal{A}}_{\boldsymbol{\gamma}}},\;
\bm\nu_{\bar{\bm{w}}},\; \sigma^2_{(\bar{\mathcal{A}})},\;
\boldsymbol{\gamma}
&\sim&
\mathcal{N}\!\big(\bm\mu_{\bar{w}}^{(\boldsymbol{\gamma})},\;
(\mathbf{\Lambda}_{\bar{w}}^{(\boldsymbol{\gamma})})^{-1}\big)
\\[6pt]
\sigma^{2}_{(0)}
&\mid&
\mathcal{D}_0,\; \bm{w}^{(\mathcal{A})},\; \bm\delta
&\sim&
\mathrm{InvGamma}\!\big(a_{0}^{*},\;
b_{0}^{*}\big)
\\[6pt]
\sigma^{2}_{(\mathcal{A})}
&\mid&
\mathcal{D}_{\mathcal{A}_{\boldsymbol{\gamma}}},\; \bm{w}^{(\mathcal{A})}
&\sim&
\mathrm{InvGamma}\!\big(a_{\mathcal{A}_{\bm\gamma}}^{*},\;
b_{\mathcal{A}_{\bm\gamma}}^{*}\big)
\\[6pt]
\sigma^{2}_{(\bar{\mathcal{A}})}
&\mid&
\mathcal{D}_{\bar{\mathcal{A}}_{\boldsymbol{\gamma}}},\; \bm{w}^{(\bar{\mathcal{A}})}
&\sim&
\mathrm{InvGamma}\!\big(a_{\bar{\mathcal{A}}_{\bm\gamma}}^{*},\;
b_{\bar{\mathcal{A}}_{\bm\gamma}}^{*}\big)
\end{array}
\]

  Update shrinkage parameters $\bm\nu_{\bm{w}}, \bm\nu_{\bar{\bm w}}, \bm\nu_{\bm\delta}$ via Gibbs, MH or AR step \;

  Compute $\bm\beta = \bm{w}^{(\mathcal{A})} + \bm\delta$\;

  \tcp*[h]{\textbf{Source Study Selection Step}}\\
\For{$k = 1,\dots,K$}{
    Propose $\gamma_k' = 1-\gamma_k$ and set 
    $\bm{\gamma}' = (\gamma_1,\dots,\gamma_{k-1},\gamma_k',\gamma_{k+1},\dots,\gamma_K)$\;
    
    Compute
    \[
    \alpha_k
    =
    \min\!\left(
    1,\;
    \frac{
    p(\bm{\gamma}' \mid \bm{\nu}, \mathcal D)
    }{
    p(\bm{\gamma} \mid \bm{\nu}, \mathcal D)
    }
    \right)
    \]
    
    Accept $\gamma_k \leftarrow \gamma_k'$ with probability $\alpha_k$\;
}

  Store $(\bm\beta^{(t)}, \bm\gamma^{(t)}, \bm\nu^{(t)}, \bm\sigma^{2(t)})$\;
}
\KwRet{$\{\bm\beta^{(t)}, \bm\gamma^{(t)}, \bm\nu^{(t)}, \bm\sigma^{2(t)}\}_{t=1}^{T},\ \left\{\sum_{t=1}^T \gamma_k^{(t)} / T\right\}_{k=1}^K$}\;
\end{algorithm}

\subsection{BLAST Horseshoe Implementation}\label{HorseshoeImplementation}
As previously noted, BLAST can be implemented with any prior from the class of global-local shrinkage priors \citep{Bahdra2016}. A prominent example is the horseshoe (HS) prior \citep{Carvalho2010TheSignals}, in which the global-local shrinkage for each coefficient is determined by $\nu_j = \lambda_j^2\tau^2$. The local shrinkage parameters,  $\lambda_j$,  govern the shrinkage of individual coefficients while the global shrinkage parameter, $\tau$, controls overall sparsity across all coefficients. The HS shrinkage prior has demonstrated robust theoretical guarantees in high-dimensional sparse settings \citep{vanderPas2014TheVectors, vanderPas2017UncertaintyDiscussion, SongLiang2023NearlyOptimalBayesianShrinkage}. 
%In a fully Bayesian framework, the HS prior is typically specified as  
%\begin{align*}
%    \lambda_j \sim \text{C}^+(0, 1) \quad\text{for } j = 1, \ldots, p, \quad
%    \tau     \sim \text{C}^+(0, 1).
%\end{align*}
%These priors make the horseshoe particularly well-suited for handling sparse vectors like the source coefficients and contrasts in our transfer learning problem. % by enabling heavy-tailed behavior to retain large signals while enforcing strong shrinkage on small coefficients through a sharp peak at zero \citep{Carvalho2009HandlingHorseshoe}. 

HS priors may be placed on the source regression coefficients and contrasts, s.t.
$$
\begin{array}{lclcl}
{\bm w}^{(\mathcal{A})} &\mid& \sigma^2_{(\mathcal{A})},\, \tau_{(\mathcal{A})},\, c,\, \Lambda_{(\mathcal{A})} &\sim& \mathcal{N}\left({\bf 0},\; \sigma^2_{(\mathcal{A})} \Lambda_{(\mathcal{A})} \tau^2_{(\mathcal{A})} \right), \\
{\bm w}^{(\bar{\mathcal{A}})} &\mid& \sigma^2_{(\bar{\mathcal{A}})},\, \tau_{(\bar{\mathcal{A}})},\, c,\, \Lambda_{(\bar{\mathcal{A}})} &\sim& \mathcal{N}\left({\bf 0},\; \sigma^2_{(\bar{\mathcal{A}})} \Lambda_{(\bar{\mathcal{A}})} \tau^2_{(\bar{\mathcal{A}})} \right), \\
\boldsymbol{\delta} &\mid& \sigma^2_{(0)},\, \tau_{(0)},\, c,\, \Lambda_{(0)} &\sim& \mathcal{N}\left({\bf 0},\; \sigma^2_{(0)} \Lambda_{(0)} \tau^2_{(0)} \right)
\end{array}
$$
where each matrix $\Lambda_{(\cdot)} = \mbox{diag}\left({\lambda}^{2}_{1(\cdot)},\cdots,{\lambda}^{2}_{p(\cdot)}\right)$, ($(\cdot)$ indexing either $\mathcal{A}$ or $\bar{\mathcal{A}}$).
Using this representation of the HS, we further assume:
$$
\begin{array}{lclclcl}
\lambda_{j(\cdot)} &\sim& \text{C}^{+}(0,1),&&
\tau_{(\cdot)}   &\sim& \text{C}^{+}(0, \psi^2).\\
\end{array}
$$
To ensure a strict notion of compatibility between informative sources and target, we further require the vector of sparse contrasts $\bm{\delta}$ to be strictly sparser than the anchoring signals ${\bm w}^{(\mathcal{A})}$. This is easily achieved by truncating the contrast global shrinkage as follows:
$$\tau_{(0)}\mid \tau_{(\mathcal{A})} \sim C^+(0,\psi^2)I(\tau_{(0)} < \tau_{(\mathcal{A})}).$$
In default analyses, the scale parameter $\psi$ is often fixed at $\psi = 1$. Alternatively, this parameter may be estimated through an empirical Bayes Gibbs sampling approach \citep{Casella2001EmpiricalSampling} (see Appendix C). Finally, the residual variance and source indicator priors are defined as follows:
$$
\begin{array}{lclclcl}
\sigma^2_{(\cdot)}  &\sim& \mathrm{IG}(\omega/2, \omega/2),&
\boldsymbol{\gamma} &\sim& f_\gamma,\\
\end{array}
$$
where typically $\omega = 1$. 

For the HS prior, there exist several efficient algorithms for sampling the regression coefficients and their associated shrinkage parameters. In particular, \citet{Johndrow2020ScalablePrior} introduce a scalable sampling algorithm that exhibits strong performance in high-dimensional settings. Appendix~C provides the implementation details for this sampling algorithm.

\section{Large Sample Behavior of BLAST}
In this section, we examine basic large-sample properties of BLAST. We show that BLAST achieves more accurate posterior concentration in the presence of informative auxiliary data. We also show that informative source selection follows standard Bayes Factor asymptotics, ensuring consistent source selection. We begin by introducing notation and asymptotic conventions used throughout.

\subsection{Notation}

We adopt standard asymptotic notation for high-dimensional regression. The number of predictors is denoted by $p = p_n$, which may grow with the sample size $n$. For a vector $\bm v \in \mathbb R^p$, we use $\|\bm v\|_0$ to denote the number of nonzero components. For any index set $\xi \subset \{1,\dots,p\}$, let $\mathbf X_\xi$ denote the submatrix of the design matrix corresponding to columns indexed by $\xi$, and let $\lambda_{\min}(\cdot)$ denote the minimum eigenvalue of a matrix.

Posterior probabilities are denoted by $\Pi(\cdot \mid \mathbf y)$, and convergence statements such as
\[
\Pi(\cdot \mid \mathbf y)
\;\xrightarrow{P_{\bm\theta^\star}^{(n)}}\; 0
\]
are interpreted as convergence in probability under the true data-generating distribution indexed by the true parameter $\bm\theta^\star$.

We use the asymptotic comparison notation $a_n \prec b_n$ to denote $a_n = o(b_n)$, and $a_n \gtrsim b_n$ to denote $a_n \ge C b_n$ for some positive constant $C$.

\subsection{Posterior Contraction under Oracle Knowledge}

Throughout this subsection, we work under the Oracle BLAST Gaussian model
in~(\ref{Aknownlik}), and adopt the required regularity assumptions from $\text{A}_1$ and $\text{A}_2$ of 
\citet{SongLiang2023NearlyOptimalBayesianShrinkage}. These assumptions are summarized below

\begin{assumption}[Design and sparsity conditions]
\label{ass:SL_conditions}
We impose the following regularity conditions, corresponding to 
Assumptions $\text{A}_1$--$\text{A}_2$ of \citet{SongLiang2023NearlyOptimalBayesianShrinkage} for a true regression coefficient vector $\bm\beta^*$.

\begin{enumerate}
    \item[(A1)] \textbf{Design conditions.}
    \begin{enumerate}
        \item[(i)] (Uniform boundedness) The covariates are uniformly bounded,
        with each column satisfying $x_j \in [-1,1]^n$ for $j=1,\dots,p_n$.
        
        \item[(ii)] (High dimensionality) The dimension satisfies $p \gtrsim n$.
        
        \item[(iii)] (Restricted eigenvalue condition) There exist an integer 
        $\bar p \succ s$ and a constant $\lambda_0 > 0$ such that for any subset 
        $\xi$ with $|\xi| \le \bar p$,
        \[
        \lambda_{\min}(X_\xi^\top X_\xi) \ge n\lambda_0.
        \]
    \end{enumerate}
    
    \item[(A2)] \textbf{Sparsity and signal strength.}
    \begin{enumerate}
        \item[(i)] (Sparsity scaling) The true regression vector 
        $\boldsymbol{\beta}^*$ is $s$-sparse and satisfies
        \[
        s \log p_n \prec n.
        \]
        
        \item[(ii)] (Signal magnitude control) The nonzero coefficients satisfy
        \[
        \max_j \left| \beta_j^* / \sigma^* \right|
        \le \gamma_3 E_n,
        \]
        for some fixed $\gamma_3 \in (0,1)$ and a sequence $E_n$
        nondecreasing in $n$.
    \end{enumerate}
\end{enumerate}
\end{assumption}

Under Assumption \ref{ass:SL_conditions}, the following theorems establish posterior contraction rates for the regression parameters in the oracle model that coincide with the minimax-optimal rates for sparse high-dimensional linear regression. 
\begin{theorem}[Posterior contraction for $\bm w$ under oracle $\mathcal A$ - known contrasts $\boldsymbol{\delta}$]
Let $\bm w^\star$ denote the true anchoring coefficients with sparsity
$s_w := \|\bm w^\star\|_0$, and define $n_w := n_0 + n_{|\mathcal A|}$ with pooled
design $\mathbf X_w := [\mathbf X^{(0)\top},\,\mathbf X^{(\mathcal A)\top}]^\top$.
Under the regularity conditions stated above, the posterior distribution of
$\bm w$ satisfies, for a sufficiently large constant $M>0$,
\[
\Pi\!\left(
\|\bm w - \bm w^\star\|_2 \ge M \varepsilon_{n_w}
\;\middle|\; \mathbf y
\right)
\;\xrightarrow{P_{\bm w^\star}^{(n_w)}}\; 0,
\]
and
\[
\Pi\!\left(
\|\bm w - \bm w^\star\|_1
\ge
M s_w^{1/2} \varepsilon_{n_w}
\;\middle|\; \mathbf y
\right)
\;\xrightarrow{P_{\bm w^\star}^{(n_w)}}\; 0,
\]
where the contraction rate is
\[
\varepsilon_{n_w}
=
\sqrt{\frac{s_w \log p}{n_w}}.
\]
\end{theorem}

An analogous result holds for the contrast parameters $\bm\delta$, as made
explicit in the following theorem.

\begin{theorem}[Posterior contraction for $\bm\delta$ under oracle $\mathcal A$ - known anchoring signals ${\bm w}^{(\mathcal{A})}$]
Let $\bm\delta^\star$ denote the true contrast vector with sparsity
$s_\delta := \|\bm\delta^\star\|_0$.
Conditional on $\bm w$, the posterior distribution of $\bm\delta$ satisfies,
for a sufficiently large constant $M>0$,
\[
\Pi\!\left(
\|\bm\delta - \bm\delta^\star\|_2 \ge M \varepsilon_{n_0}
\;\middle|\; \mathbf y, \bm w
\right)
\;\xrightarrow{P_{\bm\delta^\star}^{(n_0)}}\; 0,
\]
and
\[
\Pi\!\left(
\|\bm\delta - \bm\delta^\star\|_1
\ge
M s_\delta^{1/2} \varepsilon_{n_0}
\;\middle|\; \mathbf y, \bm w
\right)
\;\xrightarrow{P_{\bm\delta^\star}^{(n_0)}}\; 0,
\]
where
\[
\varepsilon_{n_0}
=
\sqrt{\frac{s_\delta \log p}{n_0}}.
\]
\end{theorem}

Finally, we consider a contraction result for the stacked parameter vector
$\bm\theta := (\bm w^\top,\bm\delta^\top)^\top$.

\begin{theorem}[Joint posterior contraction for $(\bm w,\bm\delta)$ under oracle $\mathcal A$]
Write the oracle BLAST model in stacked form as
\[
\mathbf y
=
\begin{pmatrix}
\mathbf y^{(0)} \\
\mathbf y^{(\mathcal A)}
\end{pmatrix}
=
\mathbf Z \bm\theta
+
\boldsymbol\varepsilon,
\qquad
\bm\theta := (\bm w^\top,\bm\delta^\top)^\top \in \mathbb R^{2p},
\]
with stacked design matrix
\[
\mathbf Z
=
\begin{pmatrix}
\mathbf X^{(0)} & \mathbf X^{(0)} \\
\mathbf X^{(\mathcal A)} & \mathbf 0
\end{pmatrix}.
\]

Let $\bm\theta^\star := ((\bm w^\star)^\top,(\bm\delta^\star)^\top)^\top$
denote the true stacked parameter, with sparsity
\[
s_\theta := \|\bm\theta^\star\|_0 = s_w + s_\delta,
\]
and let $n_w := n_0 + n_{|\mathcal A|}$ denote the total sample size in the informative set and target data.
Under the regularity conditions stated above, the posterior distribution of
$\bm\theta$ contracts at the rate
\[
\varepsilon_{n}
=
\sqrt{
\frac{s_\theta \log(2p)}{n_w}
}.
\]
Specifically, for a sufficiently large constant $M>0$,
\[
\Pi\!\left(
\|\bm\theta - \bm\theta^\star\|_2 \ge M \varepsilon_{n}
\;\middle|\; \mathbf y
\right)
\;\xrightarrow{P_{\bm\theta^\star}^{(n)}}\; 0,
\]
and
\[
\Pi\!\left(
\|\bm\theta - \bm\theta^\star\|_1
\ge
M s_\theta^{1/2} \varepsilon_{n}
\;\middle|\; \mathbf y
\right)
\;\xrightarrow{P_{\bm\theta^\star}^{(n)}}\; 0.
\]
\end{theorem}

Theorems~3.1--3.3 can be viewed as direct adaptations of the results of
\citet{SongLiang2023NearlyOptimalBayesianShrinkage}, obtained by substituting the appropriate parameter dimensions, sample sizes, and sparsity levels. The improved contraction rates, compared to a target only analysis, essentially stem from pooling informative data sources. Given the larger pooled sample, the magnitude of contraction gains depends on the level of source-target compatibility, quantified by the sparsity of the contrast vector $s_\delta$.

Importantly, the priors we consider in our transfer learning model are specified hierarchically as scale mixtures of
Gaussian distributions, and, as shown in Section~3 of
\citet{SongLiang2023NearlyOptimalBayesianShrinkage}, a broad class of
scale-mixture shrinkage priors—including the horseshoe and related global--local priors—satisfy the required prior concentration and tail conditions for the above contraction results to hold under the oracle model.

\subsection{Asymptotic Behavior of Bayes Factors for Source Selection}

We study the large-sample behavior of Bayes factors used to select informative
auxiliary studies in the BLAST framework. Model selection is conducted by comparing
marginal likelihoods corresponding to different source membership configurations
$\boldsymbol{\gamma}$, which define alternative partitions of the auxiliary studies
into informative and non-informative sets. For two candidate configurations $\boldsymbol{\gamma}^{(1)}$ and
$\boldsymbol{\gamma}^{(2)}$, model comparison is based on the Bayes factor
\[
\mathrm{BF}_{12}
\;=\;
\frac{p(\boldsymbol{\gamma}^{(1)} \mid\mathcal{D}) }
     {p(\boldsymbol{\gamma}^{(2)}\mid \mathcal{D})},
\]
where
\[
p( \boldsymbol{\gamma} \mid \mathcal{D})
=
\int p(\mathcal{D}\mid \bm\vartheta,\boldsymbol{\gamma})
     \,\pi(\bm\vartheta\mid\boldsymbol{\gamma})\,d\bm\vartheta
\]
denotes the marginal likelihood under configuration $\boldsymbol{\gamma}$,
obtained by integrating out the model parameters $\bm\vartheta$ under the
BLAST hierarchy.

In this subsection, we assume standard regularity conditions for likelihood-based
model comparison in Gaussian linear regression, including interior maximum likelihood
estimators, twice continuously differentiable log-likelihoods, nonsingular Fisher
information matrices, and priors that are continuous and strictly positive in neighborhoods
of the relevant estimators. Under these conditions, marginal likelihoods admit Laplace
approximations and Bayes factor asymptotics follow classical likelihood theory (see Appendix D).

\begin{theorem}[Bayes factor consistency for general source configurations]
\label{thm:bf_general_gamma}
Consider two source membership configurations $\boldsymbol{\gamma}^{(1)}$ and
$\boldsymbol{\gamma}^{(2)}$, defining alternative partitions of the auxiliary studies.
Let $\mathrm{BF}_{12}$ denote the Bayes factor comparing the corresponding BLAST models,
and let $n$ denote the total combined sample size of the sources. We have two possible cases:

\begin{enumerate}
\item \textbf{Non-nested configurations.}
If both configurations assign at least one study to the informative set and at least
one study to the non-informative set, then the two models have equal parameter dimension.
In this case,
\[
\log \mathrm{BF}_{12} = \Delta \ell_n + O_p(1),
\]
where $\Delta \ell_n$ denotes the difference in maximized log-likelihoods.
Moreover, there exists a constant $c$ such that
\[
\frac{1}{n}\Delta \ell_n \xrightarrow{p} c.
\]
If $\boldsymbol{\gamma}^{(1)}$ is the true configuration, then $c>0$ and
$\log \mathrm{BF}_{12} \to +\infty$ at a linear rate, implying exponential consistency
of the Bayes factor in favor of the true configuration.

\item \textbf{Boundary (nested) configurations.}
If one configuration assigns all auxiliary studies to either the informative or
non-informative set, the corresponding models are nested. Let $\bm\gamma^{(1)}$ be the larger model. In this case,
\[
\log \mathrm{BF}_{12}
=
\Delta \ell_n - \frac{r}{2}\log n + O_p(1),
\]
where $r$ denotes the difference in model dimension.
If the smaller model is true, the Bayes factor decays polynomially in $n$; if the
larger model is true, the Bayes factor grows exponentially.
\end{enumerate}
\end{theorem}

Theorem~\ref{thm:bf_general_gamma} shows that Bayes factors provide a consistent mechanism
for identifying informative auxiliary studies in BLAST. In particular, when comparing
non-nested configurations, selection is driven entirely by differences in likelihood fit,
which decompose additively across auxiliary studies, while complexity penalties arise only
in boundary cases where one configuration is nested within another. Appendix D contains further details on these results.

\section{Simulation Studies}

In this section, we conduct a series of simulations to evaluate the empirical performance of BLAST and benchmark it against existing approaches in the high-dimensional linear regression transfer learning literature. Specifically, we compare the target-only Lasso, Oracle Trans-Lasso, Trans-Lasso, $\mathcal{A}_h$-Trans-GLM, Trans-GLM, Oracle BLAST (Algorithm~\ref{algAknown}), and BLAST (Algorithm~\ref{algAunknown}). In our simulations, Oracle BLAST and BLAST consistently outperform methods relying solely on the target data, while often surpassing the performance of Trans-Lasso and Trans-GLM.

The metrics used to assess performance  included (1) Sum of Squared Estimation Errors (SSE): defined as SSE $= \sum_{j = 1}^p (\hat{\bm\beta}_j - \bm\beta_j)^2$ where $\hat{\bm\beta}_j$ is the estimate of $\bm\beta_j$; (2) Mean Squared Prediction error (MSPE): defined as $\text{MSPE}= 1/p \sum_{j = 1}^n (\hat{y}_i - y_i)^2$ where $\hat{y}$ is the predicted value of $y_i$ on a cross-validation holdout set; (3) Average width: the width of 95\% credible/confidence interval averaged over all simulations; (4) Coverage: the proportion of the 95\% credible/confidence intervals that correctly captured the true value in the simulations. Metrics (1) and (2) assess the estimation and prediction accuracy of each method, while metrics (3) and (4) evaluate the quality of uncertainty quantification provided by the corresponding intervals. We note that metrics (3) and (4) are only comparable between methods that provide interval estimates—namely, $\mathcal{A}_h$-Trans-GLM and our Bayesian methods. For completeness, we also include the target-only desparsified Lasso \citep{vandeGeer2013OnModels} as an additional point of comparison for interval estimation. 

All experiments are performed in R. The Oracle Trans-Lasso and Trans-Lasso functions are obtained from Sai Li's public GitHub repository (\url{https://github.com/saili0103/TransLasso}), and the $\mathcal{A}$-Trans-GLM and Trans-GLM methods are available in the  \texttt{glmtrans} package \citep{R-glmtrans}.

\subsection{Simulation Setup}

Our simulation setup is as follows. We set \( p = 200 \), with \( n_0 = 150 \) target samples and \( n_k = 150 \) auxiliary samples for each \( k = 1, \dots, K \) with $K = 10$. The covariates \( \mathbf{x}_i^{(k)} \) are independently drawn from a $\mathcal{N}(0, 1)$ and the error variances for target and source studies are fixed to $\sigma^2_{(0)} = \sigma^2_{(\mathcal{A})} = \sigma^2_{(\bar{\mathcal{A}})} =  1$.   
For the target parameter, we define:  
\[
\bm{\beta} = (0.5 \mathbf{1}_s, \mathbf{0}_{p-s})^T,
\]
where \( s = 6 \), meaning the first \( s \) entries are set to 0.5, and the remaining are zero.  

 We construct the source regression coefficients by introducing a structured bias to random components of the target parameter. Specifically,  for a given \( \mathcal{A} \) and $k \in \mathcal{A}$ we define:  
\[
\bm{w}_j^{(k)} = \bm{\beta} - 0.3 * \mathds{1}(j \in H_k), \quad \text{if } k \in \mathcal{A},
\]
where \( H_k \) is a random subset of \([p]\) with \( |H_k| = h \) for \( h \in \{2, 4, 6\} \). 
Similarly, for  $k\in \bar{\mathcal{A}}$ we define:  
\[
\bm{w}_j^{(k)} = \bm{\beta} - 0.5*\mathds{1}(j \in H_k), \quad \text{if } k \in \bar{\mathcal{A}},
\]  
 where \( H_k \) is a random subset of \([p]\) with \( |H_k| = 2s \).

\subsection{Simulation Results}

\subsubsection{Estimation and Prediction Accuracy}
\begin{figure}[t]
    \centering
    \includegraphics[width=1.0\linewidth]{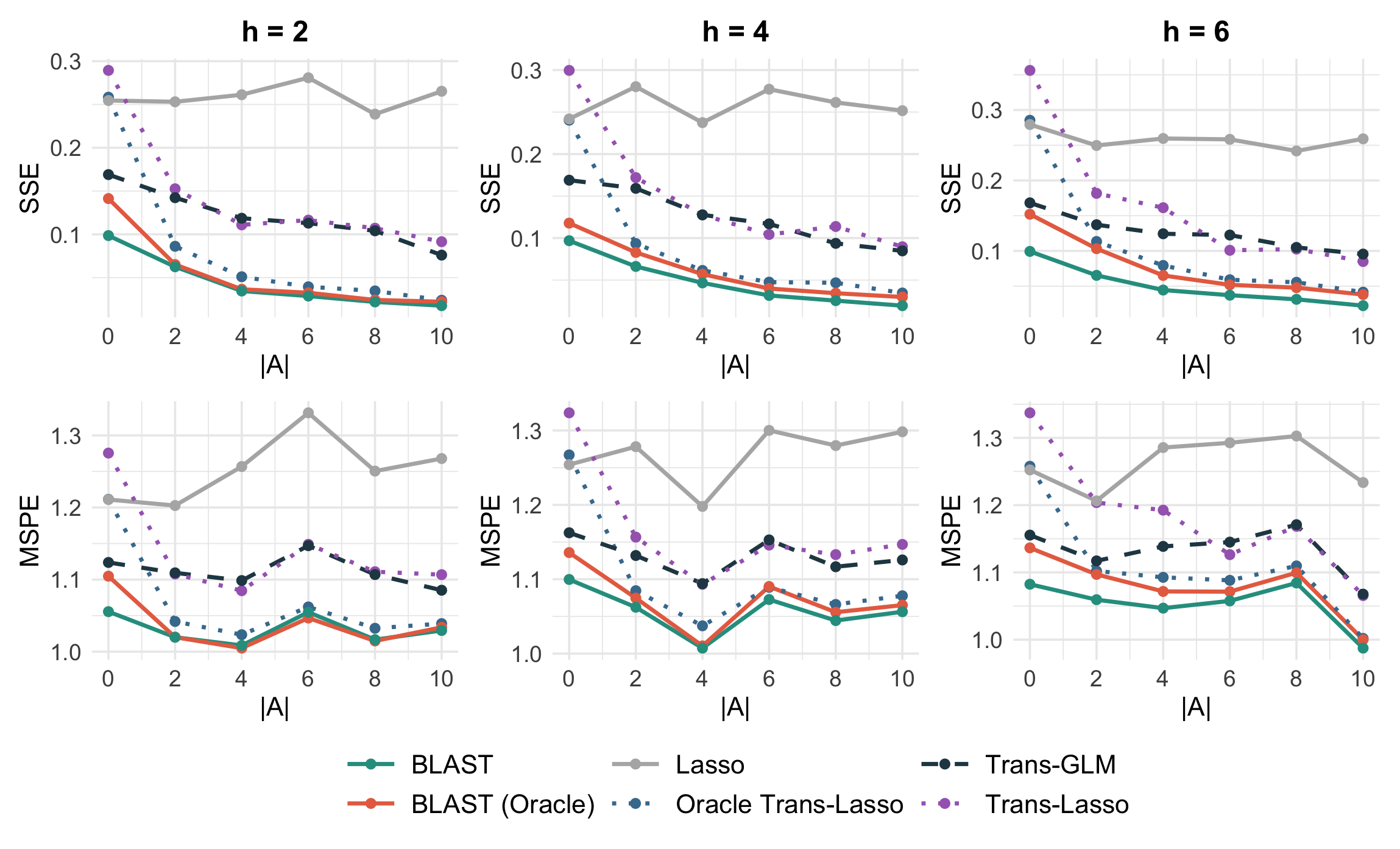}
    \caption{\small Estimation and prediction errors for various transfer learning methods with different settings of $h$ for $K = 10$. $n_k = 150$ for $k = 0,\ldots, K$, $p = 200$ and $s = 6$.  The x-axis denotes the number of informative source studies $|\mathcal{A}|$. Each point represents an average over 50 simulation replicates. 
}
    \label{fig:simulation-grid}
\end{figure}

For Oracle BLAST (Algorithm \ref{algAknown}) and BLAST (Algorithm \ref{algAunknown}), we use the HS prior formulation from Section \ref{HorseshoeImplementation}  and run $T = 3,000$ MCMC iterations with a burn-in of $1,000$ samples. For BLAST, tempering was performed in the first 90\% of burn-in samples to ensure proper mixing of the latent source inclusion indicator variables. 
Figure \ref{fig:simulation-grid} shows line plots of the average MSE of the estimator $\hat{\bm{\beta}}$ and MSPE on a holdout set as a function of the number of informative source studies. Each point in the graph is averaged over 50 independent simulations with results displayed for $h =\{2, 4, 6\}$. 

To start, we observe that estimation and prediction error of the BLAST methods tend to decrease as the number of informative source studies increases, indicating that source data are being effectively leveraged. Furthermore, BLAST and its oracle variant consistently achieve lower estimation and prediction error than competing transfer learning methods when there is a high concentration of informative data. These results appear to be largely consistent over different values of $h$. 

Comparing the BLAST methods alone, BLAST with source selection performs comparably to, and occasionally outperforms, Oracle BLAST with respect to both estimation and prediction errors. This behavior may be seen as counterintuitive since the latter method operates under knowledge of the informative set. Upon closer examination of the simulation results, we found that this discrepancy was driven primarily by the learning of sparse signals. While Oracle BLAST achieves slightly lower error for true signals, BLAST with source selection produces substantially lower mean squared error for non-signal coordinates. %\textbf{[EXPLAIN WHY?]} .

% This result may be explained by the additional regularization introduced by BLAST via its source study selection selection mechanism. By occasionally down-weighting or excluding sources, BLAST promotes sharper shrinkage of irrelevant coefficients and can therefore identify zeros more effectively than Oracle BLAST. 

\begin{figure}[t]
    \centering
    \includegraphics[width=0.85\linewidth]{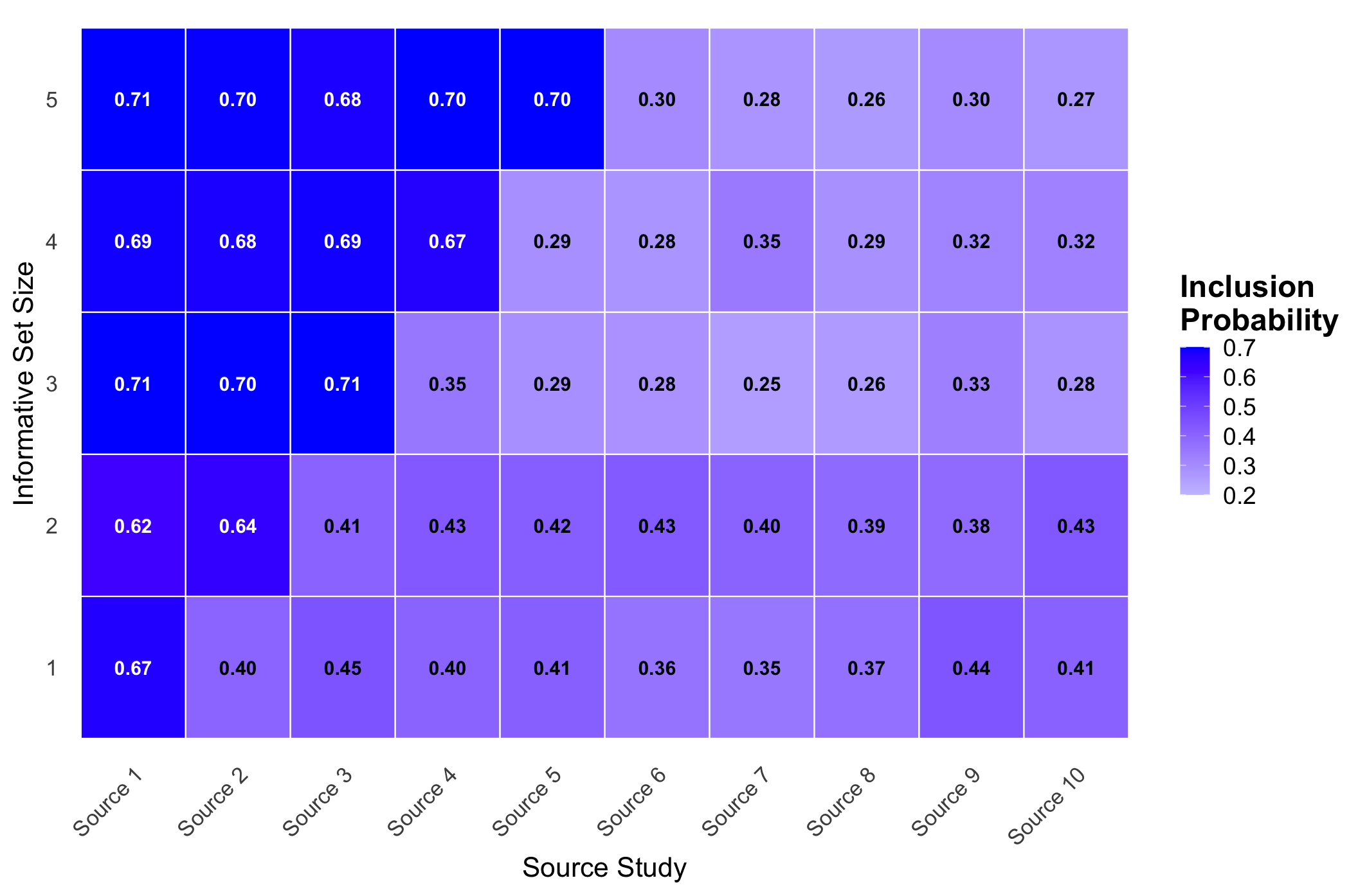}
\caption{\small
Posterior inclusion probabilities for each auxiliary study under varying informative set sizes.
Each row corresponds to a different number of truly informative source studies (from $1$ to $5$), with the informative studies always assigned to the first $|\mathcal{A}|$ positions. %The heatmap displays the posterior inclusion probability for each source study, averaged over 5{,}000 MCMC samples. 
Cells corresponding to the true informative studies are highlighted with bold white text. Prior inclusion probabilities were set to 0.5.  
}
\label{fig:sim_inclusion_heatmap}
\end{figure}

\subsubsection{Model Selection: Identification of the Informative Set of Source Datasets}

Accurate selection of informative source datasets is fundamental to avoiding negative transfer and achieving effective posterior inference for $\bm\beta$. A key feature of BLAST is its ability to learn which source studies provide useful signal for the target regression task, and to downweight those that do not. To evaluate this capability, we simulate scenarios with varying sizes of the true informative set, considering $|\mathcal{A}| \in \{1, 2, 3, 4, 5\}$ out of a total of $K = 10$ source studies and assess the proportion of posterior MCMC samples in which each source study was selected. For this simulation, to emphasize discrepancies between the target and noninformative sources, we increase the magnitude of the coefficient deviations for 
$k\in\bar{\mathcal{A}}$:
\[
\bm{w}_j^{(k)} = \bm{\beta} - 0.6*\mathds{1}(j \in H_k).
\]  

Figure~\ref{fig:sim_inclusion_heatmap} presents heatmaps of posterior inclusion probabilities for each auxiliary study, averaged over 3{,}000 posterior samples. Each row corresponds to a different informative set size, with informative studies appearing in positions $\{1,2,\ldots,|\mathcal{A}|\}$ by construction. All studies were assigned a prior inclusion probability of $0.5$.

The results demonstrate the capacity of our method to identify and leverage useful sources. BLAST effectively discriminates by assigning higher posterior inclusion probabilities to truly informative studies while reducing weight on noninformative ones. As Figure~\ref{fig:sim_inclusion_heatmap}  shows, informative studies are typically selected with posterior probabilities typically around $0.7$, whereas noninformative studies tend to be selected less often, with posterior probabilities  near or below $0.45$.

\begin{figure}[t!]
    \centering
    \includegraphics[width=0.9\linewidth]{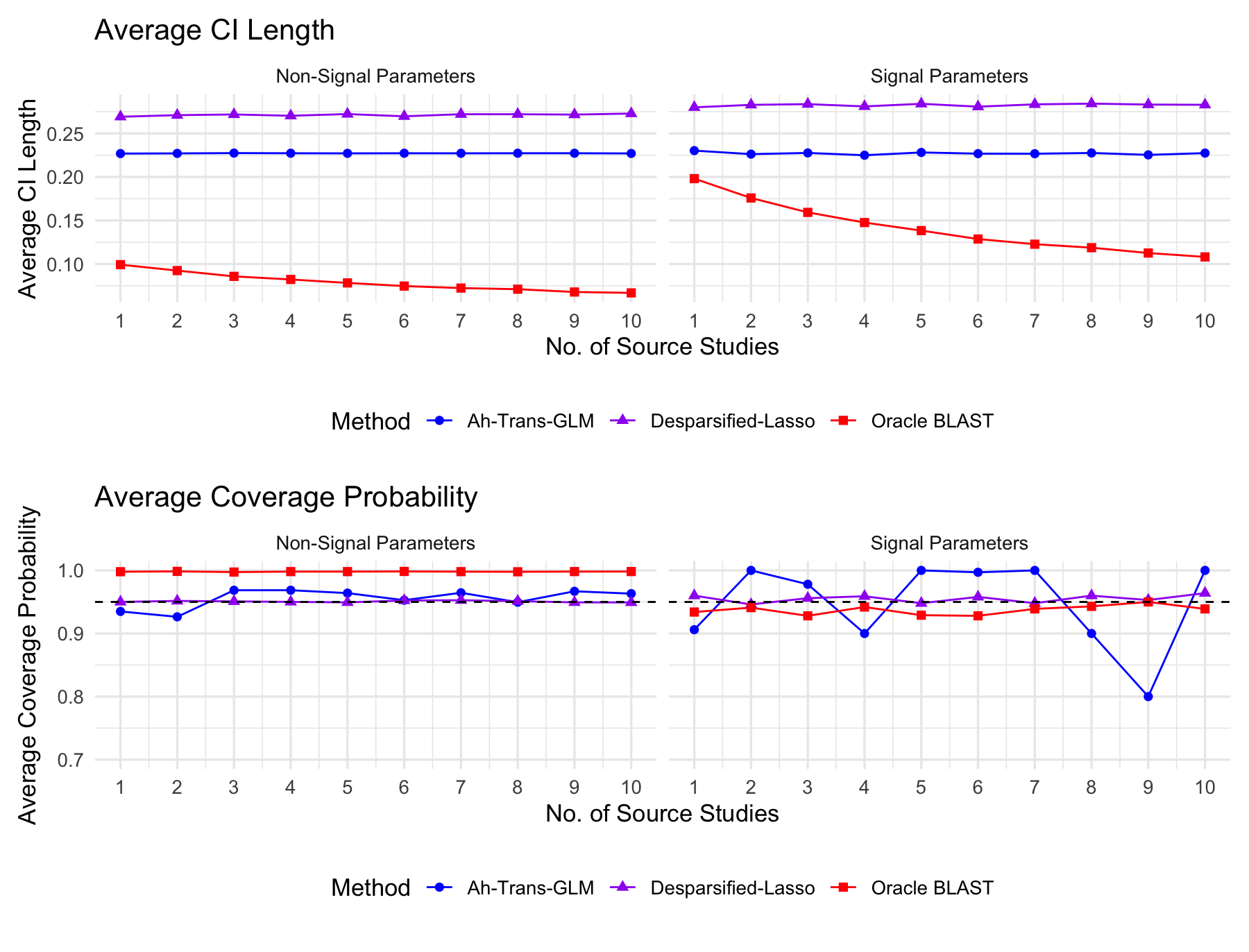}
    \caption{\small Average confidence/credible interval (CI) length (top panel) and average coverage probability (bottom panel) across varying numbers of source studies (1–10) for signal and non-signal parameters with $p = 300$ parameters and $s = 10$ signals. Results are shown for three methods: $\mathcal{A}_h$-Trans-GLM (blue), Desparsified-Lasso (purple), and Oracle BLAST (red). The dashed horizontal line in the coverage plots indicates the nominal 95\% coverage level. Each point represents an average over 50 simulation replicates.
    }
    \label{CI_oracle_plots}
\end{figure}

\subsubsection{Credible Intervals}
We compare the performance of credible or confidence intervals produced by our methods and competing methods. We start by comparing three approaches: Oracle BLAST, $\mathcal{A}_h$-Trans-GLM \citep{Tian2023TransferModels}, and the desparsified-Lasso \citep{vandeGeer2013OnModels} that produces asymptotically valid intervals using only the target data.

 In this simulation setting, we consider \( p = 300 \) predictors with \( s = 10 \) signal variables with sample sizes of $n_0 = 300$ and $n_k = 200$ for $k = 1, \ldots, K$. We vary the number of informative source studies from 1 to 10 and separate metrics for signal and non-signal parameters to better evaluate interval behavior across sparse and non-sparse dimensions. Figure~\ref{CI_oracle_plots} compares the average length and empirical coverage of the 95\% intervals produced by the three methods.

\begin{figure}[t]
    \centering
    \includegraphics[width=0.9\linewidth]{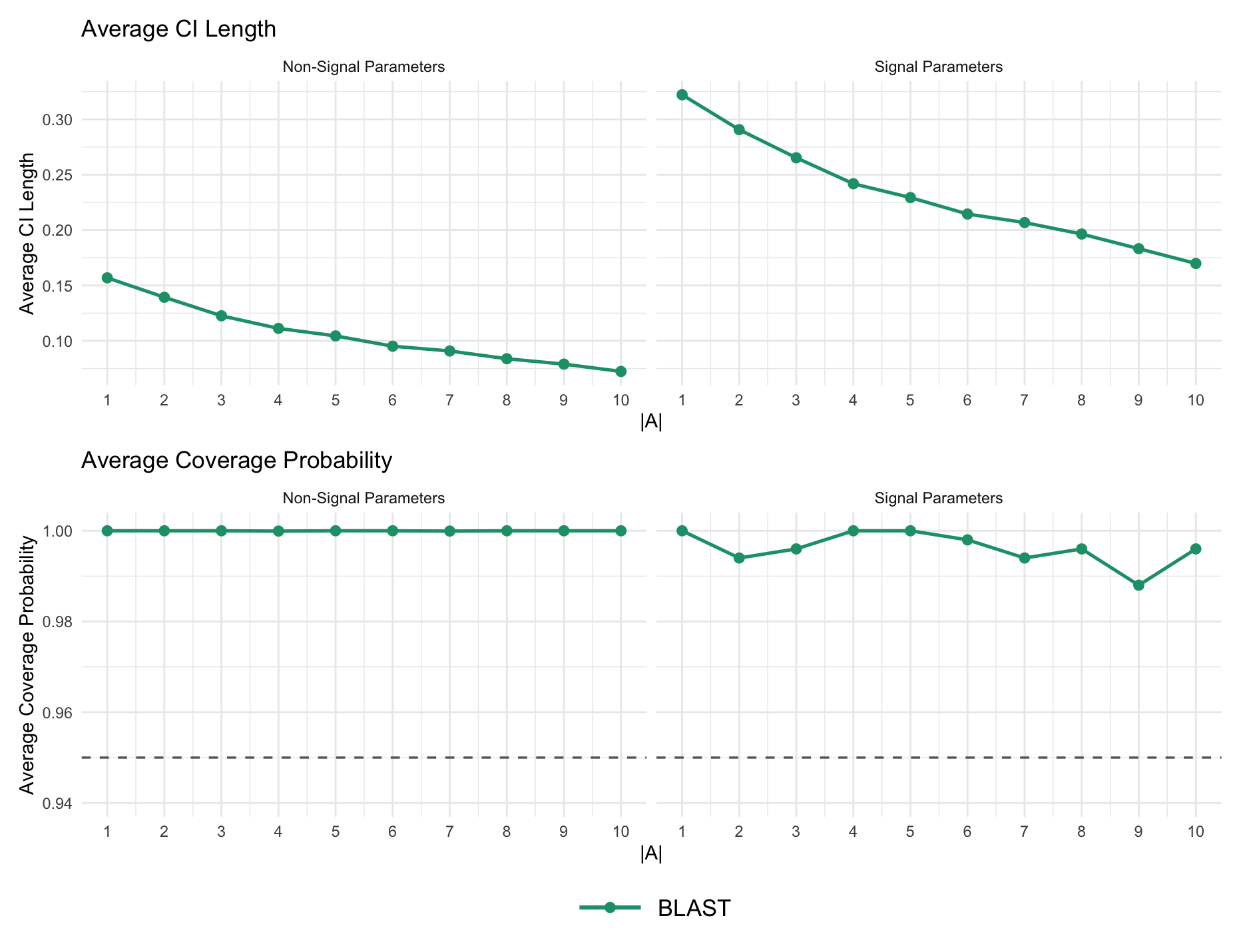}\\
    \caption{\small Average credible interval (CI) length (top panel) and average coverage probability (bottom panel) across varying numbers of informative source studies $|\mathcal{A}|$ (1–10), for signal and non-signal parameters using the BLAST method with source selection. The dashed horizontal line in the coverage plots denotes the nominal 95\% coverage level. Each point represents an average over 50 simulation replicates.
    }
    \label{fig:CI_fullBLAST_plots}
\end{figure}

For both non-signal parameters (left panels) and signal parameters (right panels), Oracle BLAST produces considerably shorter credible intervals than both competing methods while maintaining near-nominal coverage. The desparsified-Lasso and $\mathcal{A}_h$-Trans-GLM maintain mostly appropriate coverage as well. However, their interval lengths are notably longer. An overall pattern worth highlighting is the steady decrease in average credible interval length produced by Oracle BLAST as the number of informative source studies increases. This trend reflects the Bayesian model's ability to borrow strength from multiple sources to improve estimation precision. In contrast, the average interval length for the competing TL method, $\mathcal{A}_h$-Trans-GLM, remains largely flat as the number of source studies increases, indicating that its efficiency does not improve with the addition of informative studies.

We repeated this credible interval analysis using BLAST with source selection. It is worth emphasizing here that credible interval construction for BLAST accounts for uncertainty in the informative set, which is a particularly unique and attractive feature considering that interval construction for Trans-GLM conditions on an empirically determined informative set. 
Figure~\ref{fig:CI_fullBLAST_plots} shows the average length and empirical coverage probability of the credible intervals produced from BLAST  as a function of the number of informative source studies, $|\mathcal{A}|$. As $|\mathcal{A}|$ increases, we observe a pronounced decrease in average CI length for both signal and non-signal parameters, indicating greater efficiency as informative data is added. Although coverage remains slightly above the nominal 95\% level, it does not translate into a loss of interval precision, as the average CI length approaches values very close to Oracle BLAST for larger values of $|\mathcal{A}|$.

\section{Prediction of Tumor Mutational Burden} \label{sec:TCGA}

We evaluate the performance of BLAST in a real-world application involving prediction of tumor mutational burden (TMB) using molecular data from The Cancer Genome Atlas (TCGA) \citep{TheCancerGenomeAtlasResearchNetwork.2013TheProject}. TCGA provides large-scale genomic and clinical datasets across diverse cancer types, offering a natural setting for transfer learning. In this context, individual cancer types may have limited sample sizes for reliable model estimation, while related cancers may provide useful auxiliary information. This motivates the use of transfer learning methods that can selectively borrow strength across cancers. Our objective in this section is to demonstrate that BLAST improves predictive accuracy for TMB relative to naive or target-only approaches. 

\subsection{Background and Motivation and Data}
TMB represents the total number of somatic coding mutations in a tumor and has emerged as a promising biomarker for predicting immunotherapy response in cancer patients. Clinical studies have shown that high TMB is associated with better responses to immune checkpoint inhibitors and greater survival benefits in certain cancers (e.g., lung cancer and melanoma) \citep{Li2019EmergingTherapy}. While Whole Exome Sequencing (WES) provides a comprehensive and accurate measurement of TMB, it is often time-consuming and costly. As a result, several studies have explored whether sequencing specific gene panels through targeted enrichment can serve as a more practical and clinically predictive alternative to WES \citep{Fancello2019TumorChallenges, Wu2019Designingaccuracy}.

Although clinically attractive, panel-based estimates can be noisy or biased, especially since sample sizes are limited in cancer studies. In this context, transfer learning offers a powerful framework for improving TMB prediction by borrowing strength from similar cancer datasets with molecular data. By leveraging molecular profiles from multiple source studies, we can potentially enhance the predictive accuracy of a target cancer model, particularly when the sample size of the target is small or the signal is weak. Our goal is to demonstrate that incorporating information across cancers through Bayesian transfer learning leads to improved  prediction of TMB.

%\subsection{Data Analysis}

We set to evaluate the performance of our proposed method 
using
data from TCGA.  Specifically, we use the FoundationOne mRNA expression panel in a pan-cancer setting to accurately predict TMB in a target cancer. The FoundationOne gene panel has demonstrated generally reliable TMB estimation with accuracy greater than 90\% in some cancers \citep{Wu2019Designingaccuracy}. In this analysis, we consider 16 different cancers with extensive representation in peer-reviewed TCGA-based studies, though not necessarily in the context of TMB estimation. These include but are not limited to Head and Neck Squamous Cell Carcinoma, Bladder Urothelial Carcinoma, Kidney Renal Clear Cell Carcinoma, Lung Squamous Cell Carcinoma, and others. Lung Adenocarcinoma (LUAD), Lung Squamous Cell Carcinoma (LUSC), and Kidney Renal Clear Cell Carcinoma (KIRC) were chosen as target cancers due to their well-documented response to immune checkpoint inhibitors \citep{Antonia2017DurvalumabCancer, Borghaei2015NivolumabCancer, Motzer2018NivolumabCarcinoma}. The 15 TCGA cancer studies excluding the chosen target are used as source data. 

All clinical and gene expression data were extracted using the \texttt{TCGAretriever} package in R \citep{TCGAretriever}. To ensure consistency across datasets, we refined the gene panel from 324 to 303 genes by excluding those whose expression profiles were absent in at least one of the cancer studies of interest or whose expression levels were zero in more than 80\% of samples. In addition, data were standardized, and TMB was transformed to the $\log(1+\text{TMB})$ scale for all studies to reduce skewness and stabilize variance while accommodating observations with near-zero mutation counts. 

\subsection{Prediction Performance and Source Cancer Selection}
We compare the prediction performance of BLAST with Lasso, Trans-Lasso, Naive Trans-Lasso, Naive BLAST, and Trans-GLM. The BLAST implementation uses the HS prior as the chosen normal scale-mixture on all regression coefficients. Naive methods do not perform source selection and assume that all available sources are informative.

\begin{figure}[t!]
    \centering
    \begin{tabular}{c}
    \includegraphics[width=0.80\linewidth]{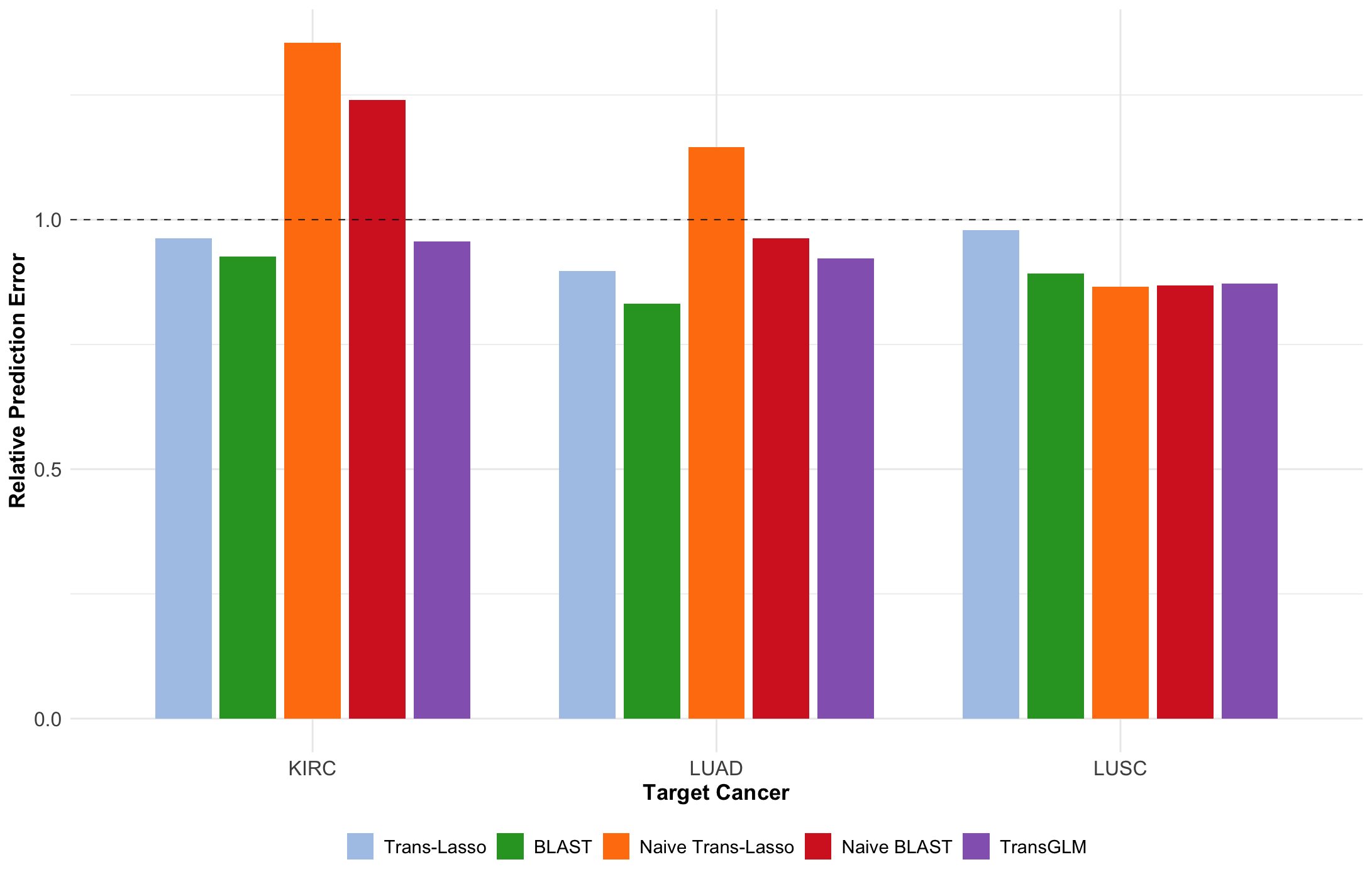}\\
     \includegraphics[width=0.85\linewidth]{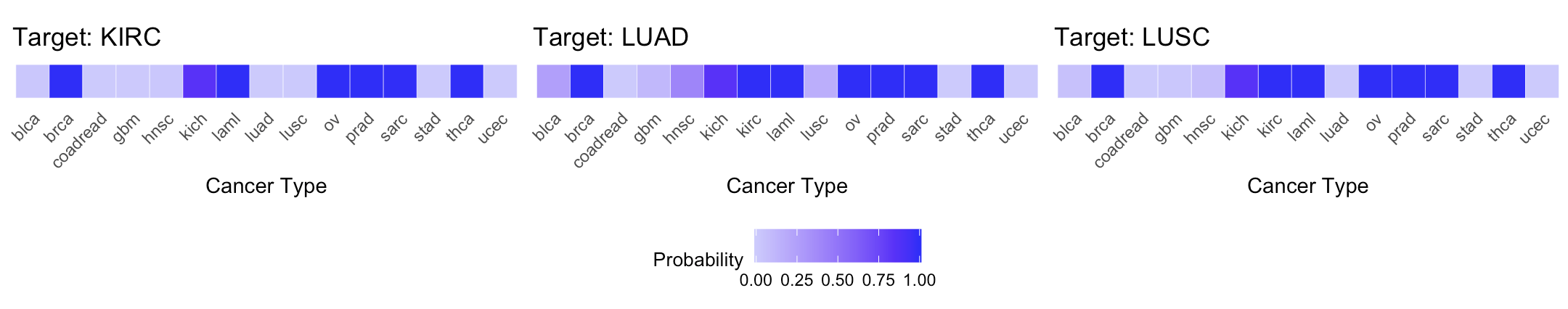}
    \end{tabular}
    \caption{\small (Top panel) Cross-validated relative prediction error %from cross-validation for transfer learning methods applied to the prediction of 
    for TMB predicted using 303 genes from the FoundationOne Gene Panel. Results are shown  
    for various cancer targets (LUAD, KIRC, LUSC) and TL methods. %as the target including lung adenocarcinoma (LUAD), kidney renal clear cell carcinoma (KIRC), and lung squamous cell carcinoma (LUSC) patients from the TCGA dataset. %Methods include both naïve and adaptive transfer learning approaches. 
    (Bottom panel) Heatmap of posterior inclusion probabilities from BLAST selection for different target cancers.}
    \label{fig:TCGA_MSPE}
\end{figure}

The target sample is split into an 80\% training set for learning of model parameters and a 20\% validation set used to assess predictive performance. Figure~\ref{fig:TCGA_MSPE} presents the cross-validated relative prediction error (RPE) for BLAST and competing transfer learning methods when predicting tumor mutational burden from the FoundationOne gene panel across different target cancers. The relative prediction error is defined as
\[
\text{RPE}_{\text{method}} =
\frac
{\text{MSPE}_{\text{method}}}{\text{MSPE}_{\text{Lasso}}},
\]
so that values less than 1 indicate improved predictive performance relative to the Lasso trained on target data alone.

Across all target cancers, transfer learning methods outperform the target-only Lasso, indicating that borrowing strength from related cancers can improve predictive accuracy. BLAST consistently achieves among the lowest relative prediction errors (RPE) across targets, with improvements of up to 17\% compared to the Lasso. In contrast, the naive approaches (Naive Trans-Lasso and Naive BLAST), which incorporate all source studies without selection, generally perform worse than their source-selection counterparts. The benefit of source selection is particularly evident for the KIRC and LUAD targets, where selectively borrowing from compatible sources helps avoid negative transfer. An exception is the LUSC target, where all TL methods perform similarly.

The heatmaps in the bottom panel of Figure~\ref{fig:TCGA_MSPE} display the corresponding posterior inclusion probabilities for each source cancer under BLAST, conditional on different targets. The posterior mass often concentrates near 0 or 1, indicating fairly decisive separation between informative and noninformative cancer sources.

\section{Summary and Discussion}

In this paper, we introduced BLAST, a novel Bayesian method for multi-source transfer learning in high-dimensional linear regression. Through the use of shrinkage priors, BLAST can robustly and adaptively learn the underlying sparsity structure and infer the regression coefficients for a target dataset via MCMC sampling effectively by leveraging information from available source data. To avoid negative transfer in cases where the informative set of source studies is unknown, BLAST incorporated a source study selection mechanism. This mechanism distinguished informative from noninformative source datasets by introducing latent source inclusion indicators that were learned in a data-driven manner using marginal likelihood evaluations. Moreover, we  established theoretical guarantees for posterior concentration and selection consistency for our method. Lastly, our empirical results for both simulated and real-world genomic datasets demonstrated that BLAST consistently outperforms the Lasso using the target data alone, and achieves comparable performance and superior uncertainty quantification compared to existing transfer learning approaches. The \texttt{BLASTreg} R package that implements our methods, along with scripts to reproduce the TCGA analysis, is available at \url{https://github.com/TelescaLab/BLASTreg}.

Future directions include extending BLAST to non-Gaussian outcomes and incorporating non-linear effects via Bayesian Gaussian processes or neural networks. Furthermore, in this paper, we did not explicitly account for source heterogeneity, which may introduce bias in posterior inference. In practice, cross-study heterogeneity may arise from covariate shifts, batch effects, different measurement platforms, or differences in study design. Explicitly modeling this heterogeneity is a natural next step to improve robustness and generalizability of the proposed method.

\phantomsection\label{supplementary-material}
\bigskip
%\newpage
\begin{center}

{\large\bf SUPPLEMENTARY MATERIAL}

\end{center}

\begin{description}
\item[Appendices:] 
 Appendix containing four sections: Appendix A (Overview of Bayesian Global-Local Shrinkage), Appendix B (General Derivations under BLAST),  Appendix C (Horseshoe Prior BLAST Implementation Details), and Appendix D (Details of Asymptotic Behavior of Bayes Factors for Source Selection).

\item[GitHub repository:] 
The GitHub repository \url{https://github.com/TelescaLab/BLASTreg} contains the \texttt{BLASTreg} R package, which implements the methods proposed in the article, as well as scripts to reproduce the TCGA dataset used in Section~\ref{sec:TCGA}.
\end{description}

\bibliography{references.bib}

\newpage

\appendix

\end{document}

\begin{center}
{\Large \textbf{Bayesian Transfer Learning for High-Dimensional Linear Regression via Adaptive Shrinkage}}\\[4pt]
{\Large \textbf{Supplementary Materials}}\\[6pt]
{\normalsize Parsa Jamshidian and Donatello Telesca}\\
{\normalsize Department of Biostatistics, University of California, Los Angeles}
\end{center}

\vspace{0.5cm}
\begin{center}
{\Large \textbf{Appendix}}
\end{center}
\numberwithin{equation}{section}

\section{Overview of Bayesian Global-Local Shrinkage}\label{sec:BayesShrink}
Our work is based on continuous shrinkage priors as a mechanism to introduce approximate sparsity through Gaussian scale mixtures \citep{PolsonScott2011, Johndrow2020ScalablePrior}. Let \( \mathbf{X} \in \mathbb{R}^{n\times p} \) be the design matrix and \( \mathbf{y} \in \mathbb{R}^n \) be the observed outcome vector. Shrinkage regression relies on Normal sampling, s.t. 
\begin{align}
    \mathbf{y} \mid \mathbf{X}, \bm{\theta}, \sigma^2 \sim \mathcal{N}(\mathbf{X}\bm{\theta}, \sigma^2 \mathbf{I}),
\end{align}
where \( \bm{\theta} \in \mathbb{R}^p \) are the regression coefficients and \( \mathbf{I} \) is the identity matrix.  A common shrinkage prior is represented as independent scale mixtures of normals,  such that:
%\begin{equation}\label{shrinkage_prior_normal}
%\begin{array}{lcll}
%     \theta_j \mid \sigma^2, \nu_j &\sim& \mathcal{N}(0, \sigma^2 \nu_j), & \text{for } j = 1, %\ldots, p,\\
%     \nu_j &\sim& f, \\
%     \sigma &\sim& g; 
%\end{array}
%\end{equation}
\begin{equation}\label{shrinkage_prior_normal}
\theta_j \mid \sigma^2, \nu_j \sim \mathcal{N}(0, \sigma^2 \nu_j),  \;\;\nu_j \sim f,   \;\;\sigma \sim g, \;\;  (\text{for } j = 1, \ldots, p), 
\end{equation}
where \( \nu_j \) represents a local scale parameter,  that governs the degree of shrinkage applied to \( \theta_j \), and $f$ and $g$ are densities with support  $(0,\infty)$. The density $f$ is often referred to as a \emph{mixing distribution}. Table \ref{tab:nu_options} summarizes several commonly used specifications of $\nu_j$ giving rise to different shrinkage priors. As shown in the table, different parametrizations and choices of mixing distributions for $\nu_j$ lead to distinct shrinkage behaviors, ranging from global regularization to coefficient-specific shrinkage. 

 \begin{table}[!h]
\centering
\begin{threeparttable}
\caption{\small Common specifications of shrinkage components $\nu_j$ in $\theta_j \mid \sigma^2, \nu_j \sim \mathcal{N}(0,\sigma^2 \nu_j)$.}
\label{tab:nu_options}
\footnotesize
\begin{tabular}{@{} p{3.2cm} p{5.3cm} p{4cm} @{}}
\toprule
Model / Prior & Variance component $\nu_j$ & Reference \\
\midrule
Ridge (Gaussian) 
& $\nu_j = \tau^2$, with $\tau$ fixed or $\tau \sim f$ 
& \citet{Hsiang1975ARegression} \\[.1in]
Bayesian Lasso 
& $\nu_j = \tau_j$, with $\tau_j \sim \text{Exp}(\lambda^2/2),$ 
& \citet{Park2008TheLasso} \\
 
& $\lambda^2 \sim \text{Gamma}(r,\delta)$ 
&  \\[.1in]
Bayesian Elastic Net
&
$\nu_j = \dfrac{1}{\lambda_2}\dfrac{\tau_j-1}{\tau_j},$
\newline
$\tau_j \sim \mathrm{TG}\!\left(\tfrac12,\dfrac{8\lambda_2\sigma^2}{\lambda_1^2},(1,\infty)\right)$
&
\citet{Li2010TheNet} \\[.05in]

Horseshoe 
& $\nu_j = \lambda_j^{2}\tau^{2}$, with $\lambda_j \sim \text{C}^+(0,1),$ 
& \citet{Carvalho2010TheSignals} \\
& $\tau \sim \text{C}^+(0,1)$ &  \\
\bottomrule
\end{tabular}
\begin{tablenotes}[flushleft]
\footnotesize
\item Note: $\text{C}^+(0,1)$ denotes the standard half-Cauchy distribution on $\mathbb{R}_+$. $\mathrm{TG}(a,b,(c,\infty))$ denotes a Gamma$(a,b)$ distribution
truncated to the interval $(c,\infty)$.

\end{tablenotes}
\end{threeparttable}
\end{table}

Among the many approaches used to induce sparsity in high-dimensional regression, continuous shrinkage priors formulated through Gaussian scale mixtures are particularly attractive due to their flexibility and compatibility with hierarchical Bayesian modeling. From a flexibility standpoint, the presence of coefficient-specific local scale parameters enables adaptive shrinkage, allowing strong signals to remain largely unshrunk while aggressively regularizing noise-level coefficients. As for computational compatibility, the scale mixture representation in (\ref{shrinkage_prior_normal}) facilitates, in principle, efficient sampling via Markov chain Monte Carlo (MCMC) methods such as Gibbs sampling or Hamiltonian Monte Carlo (HMC) \citep{Bhattacharya2016FastRegression, Johndrow2020ScalablePrior, JinTan2021FastMCMC}.
Our goal is to leverage Bayesian shrinkage in the high-dimensional multi-source transfer learning setting to achieve both stable estimation and principled uncertainty quantification. Shrinkage priors regularize high-dimensional regression coefficients while maintaining the flexibility to capture strong signals, and Bayesian posterior inference offers a coherent framework for quantifying uncertainty, which is crucial when target data are limited.

\section{General Derivations under BLAST}

\subsection{Full Conditional Distributions}
\label{app:fullconds}

We collect the full conditional distributions used in the Gibbs samplers for both the
$\mathcal A$-known (Oracle BLAST) and $\mathcal A$-unknown (BLAST) models.

\paragraph{Notation.}
Define the block-specific local shrinkage vectors
\[
\boldsymbol\nu^{\bm w} = (\nu^{\bm w}_1,\ldots,\nu^{\bm w}_p)^\top,\quad
\boldsymbol\nu^{\bm \delta} = (\nu^{\bm \delta}_1,\ldots,\nu^{\bm \delta}_p)^\top,\quad
\boldsymbol\nu^{\bm w^{\bar{\mathcal A}}}
= (\nu^{\bm w^{\bar{\mathcal A}}}_1,\ldots,\nu^{\bm w^{\bar{\mathcal A}}}_p)^\top,
\]
and the associated diagonal matrices
\[
\mathbf{D}_{(\mathcal A)}=\operatorname{diag}(\boldsymbol\nu^{\bm w}),\quad
\mathbf{D}_{(0)}=\operatorname{diag}(\boldsymbol\nu^{\bm \delta}),\quad
\mathbf{D}_{(\bar{\mathcal A})}=\operatorname{diag}(\boldsymbol\nu^{\bm w^{\bar{\mathcal A}}}).
\]
We write $\mathrm{IG}(a,b)$ for the inverse-gamma distribution with shape $a$ and scale $b$,
and $\|\cdot\|$ for the Euclidean norm.

\subsubsection{Oracle BLAST: $\mathcal A$-known model}

\paragraph{Model.}
\[
\begin{aligned}
\mathbf{y}^{(\mathcal A)} \mid \bm w^{(\mathcal A)},\sigma^2_{(\mathcal A)} 
&\sim \mathcal N\!\big(\mathbf{X}^{(\mathcal A)}\bm w^{(\mathcal A)},\,
\sigma^2_{(\mathcal A)} I\big),\\
\mathbf{y}^{(0)} \mid \bm w^{(\mathcal A)},\boldsymbol\delta,\sigma^2_{(0)} 
&\sim \mathcal N\!\big(\mathbf{X}^{(0)}(\bm w^{(\mathcal A)}+\boldsymbol\delta),\,
\sigma^2_{(0)} I\big),
\end{aligned}
\]
with priors
\[
\bm w^{(\mathcal A)}\sim \mathcal N\!\big(\mathbf{0},\sigma^2_{(\mathcal A)}\mathbf{D}_{(\mathcal A)}\big),
\qquad
\boldsymbol\delta\sim \mathcal N\!\big(\mathbf{0},\sigma^2_{(0)}\mathbf{D}_{(0)}\big).
\]

\paragraph{Full conditional for $\bm w^{(\mathcal A)}$.}
Define
\[
\mathbf{\Lambda}_w
=
\frac{1}{\sigma^2_{(\mathcal A)}}
\Big((\mathbf{X}^{(\mathcal A)})^\top\mathbf{X}^{(\mathcal A)}+\mathbf{D}_{(\mathcal A)}^{-1}\Big)
+
\frac{1}{\sigma^2_{(0)}}(\mathbf{X}^{(0)})^\top\mathbf{X}^{(0)},
\]
\[
\bm\mu_w
=
\mathbf{\Lambda}_w^{-1}
\!\left\{
\frac{1}{\sigma^2_{(\mathcal A)}}(\mathbf{X}^{(\mathcal A)})^\top\mathbf{y}^{(\mathcal A)}
+
\frac{1}{\sigma^2_{(0)}}(\mathbf{X}^{(0)})^\top
\big(\mathbf{y}^{(0)}-\mathbf{X}^{(0)}\boldsymbol\delta\big)
\right\}.
\]
Then
\[
\bm w^{(\mathcal A)} \mid \text{rest}
\sim \mathcal N\!\big(\bm\mu_w,\;\mathbf{\Lambda}_w^{-1}\big).
\]

\paragraph{Full conditional for $\boldsymbol\delta$.}
Let
\[
\mathbf{\Lambda}_\delta
=
\frac{1}{\sigma^2_{(0)}}
\Big((\mathbf{X}^{(0)})^\top\mathbf{X}^{(0)}+\mathbf{D}_{(0)}^{-1}\Big),
\qquad
\bm\mu_\delta
=
\mathbf{\Lambda}_\delta^{-1}
\!\left\{
\frac{1}{\sigma^2_{(0)}}(\mathbf{X}^{(0)})^\top
\big(\mathbf{y}^{(0)}-\mathbf{X}^{(0)}\bm w^{(\mathcal A)}\big)
\right\}.
\]
Then
\[
\boldsymbol\delta \mid \text{rest}
\sim \mathcal N\!\big(\bm\mu_\delta,\;\mathbf{\Lambda}_\delta^{-1}\big).
\]

\paragraph{Full conditionals for variance parameters.}
With independent priors
$\sigma^2_{(\mathcal A)}\sim \mathrm{IG}(a_{\mathcal A},b_{\mathcal A})$
and
$\sigma^2_{(0)}\sim \mathrm{IG}(a_{0},b_{0})$,
\[
\sigma^2_{(\mathcal A)} \mid \text{rest}
\sim \mathrm{IG}\!\left(
a_{\mathcal A}+\frac{n_{\mathcal A}+p}{2},\;
b_{\mathcal A}+
\frac{\|\mathbf{y}^{(\mathcal A)}-\mathbf{X}^{(\mathcal A)}\bm w^{(\mathcal A)}\|^2
+ (\bm w^{(\mathcal A)})^\top \mathbf{D}_{(\mathcal A)}^{-1}\bm w^{(\mathcal A)}}{2}
\right),
\]
\[
\sigma^2_{(0)} \mid \text{rest}
\sim \mathrm{IG}\!\left(
a_{0}+\frac{n_{0}+p}{2},\;
b_{0}+
\frac{\|\mathbf{y}^{(0)}-\mathbf{X}^{(0)}(\bm w^{(\mathcal A)}+\boldsymbol\delta)\|^2
+ \boldsymbol\delta^\top \mathbf{D}_{(0)}^{-1}\boldsymbol\delta}{2}
\right).
\]

\subsubsection{BLAST: $\mathcal A$-unknown model}

The $\mathcal A$-unknown model augments the Oracle BLAST specification by introducing
a noninformative-source block $\bm w^{(\bar{\mathcal A})}$ indexed by the inclusion vector
$\bm\gamma$.

\paragraph{Model.}
\[
\begin{aligned}
\mathbf{y}^{(\mathcal{A}_{\bm\gamma})} \mid \bm w^{(\mathcal A)},\sigma^2_{(\mathcal A)}
&\sim \mathcal N\!\big(\mathbf{X}^{(\mathcal{A}_{\bm\gamma})}\bm w^{(\mathcal A)},\,
\sigma^2_{(\mathcal A)}I\big),\\
\mathbf{y}^{(0)} \mid \bm w^{(\mathcal A)},\boldsymbol\delta,\sigma^2_{(0)}
&\sim \mathcal N\!\big(\mathbf{X}^{(0)}(\bm w^{(\mathcal A)}+\boldsymbol\delta),\,
\sigma^2_{(0)}I\big),\\
\mathbf{y}^{(\bar{\mathcal{A}}_{\bm\gamma})} \mid \bm w^{(\bar{\mathcal A})},\sigma^2_{(\bar{\mathcal A})}
&\sim \mathcal N\!\big(\mathbf{X}^{(\bar{\mathcal{A}}_{\bm\gamma})}\bm w^{(\bar{\mathcal A})},\,
\sigma^2_{(\bar{\mathcal A})}I\big),
\end{aligned}
\]
with priors
\[
\bm w^{(\mathcal A)}\sim \mathcal N\!\big(0,\sigma^2_{(\mathcal A)}\mathbf{D}_{(\mathcal A)}\big),\quad
\boldsymbol\delta\sim \mathcal N\!\big(0,\sigma^2_{(0)}\mathbf{D}_{(0)}\big),\quad
\bm w^{(\bar{\mathcal A})}\sim \mathcal N\!\big(0,\sigma^2_{(\bar{\mathcal A})}\mathbf{D}_{(\bar{\mathcal A})}\big).
\]
The full conditional for $\boldsymbol\delta$ is identical to that of the $\mathcal A$-known case.
The remaining conditionals are given below.

\paragraph{Full conditional for $\bm w^{(\mathcal A)}$.}
Replace $(\mathbf{X}^{(\mathcal A)},\mathbf{y}^{(\mathcal A)})$ by
$(\mathbf{X}^{(\mathcal{A}_{\bm\gamma})},\mathbf{y}^{(\mathcal{A}_{\bm\gamma})})$
in the Oracle BLAST expressions to obtain
\[
\bm w^{(\mathcal A)} \mid \text{rest},\bm\gamma
\sim \mathcal N\!\big(\bm\mu_w^{(\gamma)},\;(\mathbf{\Lambda}_w^{(\gamma)})^{-1}\big).
\]

\paragraph{Full conditional for $\bm w^{(\bar{\mathcal A})}$.}
Define
\[
\mathbf{\Lambda}_{\bar w}^{(\gamma)}
=
\frac{1}{\sigma^2_{(\bar{\mathcal A})}}
\Big((\mathbf{X}^{(\bar{\mathcal A}_{\bm\gamma})})^\top
\mathbf{X}^{(\bar{\mathcal A}_{\bm\gamma})}
+
\mathbf{D}_{(\bar{\mathcal A})}^{-1}\Big),
\]
\[
\bm\mu_{\bar w}^{(\gamma)}
=
(\mathbf{\Lambda}_{\bar w}^{(\gamma)})^{-1}
\left\{
\frac{1}{\sigma^2_{(\bar{\mathcal A})}}
(\mathbf{X}^{(\bar{\mathcal A}_{\bm\gamma})})^\top
\mathbf{y}^{(\bar{\mathcal A}_{\bm\gamma})}
\right\}.
\]
Then
\[
\bm w^{(\bar{\mathcal A})} \mid \text{rest},\bm\gamma
\sim \mathcal N\!\big(\bm\mu_{\bar w}^{(\gamma)},\;
(\mathbf{\Lambda}_{\bar w}^{(\gamma)})^{-1}\big).
\]

\paragraph{Full conditionals for variance parameters.}
With priors
$\sigma^2_{(\mathcal A)}\sim \mathrm{IG}(a_{\mathcal A},b_{\mathcal A})$,
$\sigma^2_{(0)}\sim \mathrm{IG}(a_{0},b_{0})$, and
$\sigma^2_{(\bar{\mathcal A})}\sim \mathrm{IG}(a_{\bar{\mathcal A}},b_{\bar{\mathcal A}})$,
the corresponding inverse-gamma full conditionals follow by replacing the
$\mathcal A$-known sufficient statistics with their $\bm\gamma$-indexed counterparts.

\paragraph{Shrinkage parameters and inclusion indicators.}
Local scale parameters in $\mathbf{D}_{(\mathcal A)}$, $\mathbf{D}_{(0)}$, and $\mathbf{D}_{(\bar{\mathcal A})}$
are updated via univariate steps: Gibbs updates when a conjugate representation is available,
and otherwise generic accept--reject or Metropolis--Hastings updates.
Inclusion indicators $\bm\gamma$ are updated through an MH step using the marginal likelihood calculations
described in Appendix~\ref{app:marg_lik_calc}.

\subsection{The Source Study Selection Step: Computational Considerations}
\label{app:marg_lik_calc}

\subsubsection{Expression for the Marginal Likelihood}
Let $\boldsymbol{\theta}$ denote the full vector of regression parameters, 
$\mathbf{y}$ the collection of target and source outcomes, and 
$\mathbf{X}$ the associated design matrices. Conditional on the shrinkage 
hyperparameters $\boldsymbol{\nu}$ and a fixed configuration $\boldsymbol{\gamma}$, 
the marginal likelihood is
\begin{equation}\label{marginal_likelihood_horseshoe}
p(\mathbf{y} \mid \mathbf{X}, \boldsymbol{\nu}, \boldsymbol{\gamma})
= \int 
p(\mathbf{y} \mid \mathbf{X}, \boldsymbol{\theta}, \boldsymbol{\sigma}^2)\,
p(\boldsymbol{\theta} \mid \boldsymbol{\sigma}^2, \boldsymbol{\nu})\,
p(\boldsymbol{\sigma}^2)\,
d\boldsymbol{\theta}\, d\boldsymbol{\sigma}^2,
\end{equation}
which forms the key quantity for computing $p(\bm\gamma \mid\bm\nu, \mathcal{D})$ in the Metropolis--Hastings acceptance ratio. This marginal likelihood serves as the Bayesian model evidence for a given configuration of $\boldsymbol{\gamma}$, quantifying how well that choice of informative sources explains the observed data under the specified model. Conveniently, conditional on the shrinkage parameters $\bm\nu$, the marginal likelihood admits a closed-form expression for shrinkage priors of the form in \citet{Bahdra2016}. We briefly detail its computation below.

Define 
\[
\bm\theta = \left(\boldsymbol{\delta}^{\top},\; (\bm{w}^{(\mathcal{A})})^{\top},\; (\bm{w}^{(\bar{\mathcal{A}})})^{\top}\right)^{\top},
\]
\[
\mathbf{y} = \left(\mathbf{y}^{(0)\top},\; \mathbf{y}^{(\mathcal{A})\top},\; \mathbf{y}^{(\bar{\mathcal{A}})\top}\right)^{\top}.
\]
Let $\bm\nu = \big((\boldsymbol\nu_{\bm w})^{\top},\,(\boldsymbol\nu_{\bm \delta})^{\top},\,(\boldsymbol\nu_{\bm w^{\bar{\mathcal A}}})^{\top}\big)^{\top}$
collect all local shrinkage parameters, with associated diagonal matrices
\[
\mathbf{D}_{(\mathcal A)}:=\operatorname{diag}(\boldsymbol\nu_{\bm w}),\qquad
\mathbf{D}_{(0)}:=\operatorname{diag}(\boldsymbol\nu_{\bm \delta}),\qquad
D_{(\bar{\mathcal A})}:=\operatorname{diag}(\boldsymbol\nu_{\bm w^{\bar{\mathcal A}}}).
\]
Finally, for computational convenience in the calculations, we assume a common residual variance between the source and target
\[
\sigma^2 = \sigma^2_{(0)}=\sigma^2_{(\mathcal A)}.
\]
Under the conditional independence of the informative+target block and the noninformative block for a given source configuration $\bm\gamma$,
this marginal likelihood factorizes as
\begin{align*}
p(\mathbf{y}\mid \mathbf{X},\bm\nu, \bm\gamma)
&=
p\!\left(\mathbf{y}^{(0)},\mathbf{y}^{(\mathcal A)} \mid \mathbf{X}^{(0)},\mathbf{X}^{(\mathcal A)},\bm\nu\right)
\;\times\;
p\!\left(\mathbf{y}^{(\bar{\mathcal A})}\mid \mathbf{X}^{(\bar{\mathcal A})},\bm\nu\right).
\end{align*} 
Marginalizing over $\bm w^{(\bar{\mathcal A})}$ and $\sigma^2$ under an $\mathrm{IG}(1/2,1/2)$ prior yields
\begin{align*}
p(\mathbf{y}^{(\bar{\mathcal A})}\mid \bm\nu)
&= (2\pi)^{-\bar{n}/2}\,|\mathbf{D}_{(\bar{\mathcal A})}|^{-1/2}\,
\left| (\mathbf{X}^{(\bar{\mathcal A})})^\top \mathbf{X}^{(\bar{\mathcal A})} + \mathbf{D}_{(\bar{\mathcal A})}^{-1} \right|^{-1/2} \\
&\quad \times \Gamma\!\left( \tfrac{\bar{n}+p-1}{2} \right)
\left[\tfrac{ \mathbf{y}^{(\bar{\mathcal A})\top} (\mathbf{y}^{(\bar{\mathcal A})} - \widehat{\mathbf{y}}_{(\bar{\mathcal A})}) + 1}{2}\right]^{-(\bar{n}+p-1)/2},
\end{align*}
where
\[
\widehat{\mathbf{y}}_{(\bar{\mathcal A})}
=
\mathbf{X}^{(\bar{\mathcal A})}
\Big((\mathbf{X}^{(\bar{\mathcal A})})^\top \mathbf{X}^{(\bar{\mathcal A})} + \mathbf{D}_{(\bar{\mathcal A})}^{-1}\Big)^{-1}
(\mathbf{X}^{(\bar{\mathcal A})})^\top \mathbf{y}^{(\bar{\mathcal A})}.
\]
Let
\[
\mathbf{y}^{(\dagger)}=
\begin{bmatrix}
\mathbf{y}^{(0)}\\
\mathbf{y}^{(\mathcal A)}
\end{bmatrix},
\qquad
\mathbf{Z}=
\begin{bmatrix}
\mathbf{X}^{(0)} & \mathbf{X}^{(0)}\\
\mathbf{0}       & \mathbf{X}^{(\mathcal A)}
\end{bmatrix},
\qquad
\bm\theta_{0\mathcal A}=
\begin{bmatrix}
\boldsymbol\delta\\
\bm w^{(\mathcal A)}
\end{bmatrix}.
\]
Then the joint likelihood for $(\mathbf{y}^{(0)},\mathbf{y}^{(\mathcal A)})$ can be written as
$$\mathbf{y}^{(\dagger)}\mid \bm\theta_{0\mathcal A},\sigma^2 \sim \mathcal N(\mathbf{Z}\bm\theta_{0\mathcal A},\sigma^2 \mathbf{I}),$$
with Gaussian prior
\[
\bm\theta_{0\mathcal A}\mid \sigma^2,\bm\nu
\sim
\mathcal N\!\left(
\mathbf 0,\;
\sigma^2\,
\operatorname{diag}\!\big(\mathbf{D}_{(0)},\mathbf{D}_{(\mathcal A)}\big)
\right).
\]
Integrating out $\bm\theta_{0\mathcal A}$ and $\sigma^2$ yields a closed-form expression of the form
\begin{align*}
p(\mathbf{y}^{(0)},\mathbf{y}^{(\mathcal A)}\mid \bm\nu)
&=
(2\pi)^{-n_\dagger/2}\,
\left|\operatorname{diag}\!\big(\mathbf{D}_{(0)},\mathbf{D}_{(\mathcal A)}\big)\right|^{-1/2}\,
\left|\mathbf{Z}^\top \mathbf{Z} + \operatorname{diag}\!\big(\mathbf{D}_{(0)},\mathbf{D}_{(\mathcal A)}\big)^{-1}\right|^{-1/2} \\
&\quad \times
\Gamma\!\left(\tfrac{n_\dagger + d - 1}{2}\right)
\left[\tfrac{\mathbf{y}^{(\dagger)\top}\!\left(\mathbf{y}^{(\dagger)}-\widehat{\mathbf{y}}^{(\dagger)}\right)+1}{2}\right]^{-(n_\dagger+d-1)/2},
\end{align*}
where $n_\dagger=n_0+n_{\mathcal A}$,
$d=\dim(\bm\theta_{0\mathcal A})=2p$,
and
\[
\widehat{\mathbf{y}}^{(\dagger)}
=
\mathbf{Z}
\Big(\mathbf{Z}^\top \mathbf{Z}+\operatorname{diag}\!\big(\mathbf{D}_{(0)},\mathbf{D}_{(\mathcal A)}\big)^{-1}\Big)^{-1}
\mathbf{Z}^\top \mathbf{y}^{(\dagger)}.
\]
The posterior mass of a given source configuration 
$\bm\gamma$ can then be computed (up to a normalizing constant) as
\begin{equation}\label{marginallikelihoodcalc}
\begin{aligned}
   p(\bm\gamma \mid  \bm\nu, \mathcal{D}) 
\;\propto\;
p(\mathbf{y} \mid \mathbf{X}, \boldsymbol{\nu}, \bm\gamma)\;
f_\gamma(\bm\gamma \mid \pi),
\end{aligned}
\end{equation}
where $f_\gamma$ is the prior distribution.

\subsubsection{Efficient Computation of the Marginal Likelihood}

 Under the normal scale-mixture representation,
\[
\theta_j \mid \nu_j \sim \mathcal{N}(0,\sigma^2\nu_j),
\]
so that small values of \(\nu_j\) correspond to strong shrinkage toward zero. In
high-dimensional regimes (\(p \gg n\)), posterior realizations of global--local shrinkage
priors typically exhibit \emph{effective sparsity}: many coordinates satisfy
\(\nu_j \approx 0\) and therefore contribute negligibly to the likelihood. Direct evaluation
of the marginal likelihood using all \(p\) predictors requires repeated Cholesky
factorizations of matrices of size \(p\times p\) (or \(2p\times2p\)), resulting in
\(\mathcal{O}(p^3)\) computational cost per MCMC iteration even though only a small subset
of predictors meaningfully influence the integrand.

To exploit this structure, we evaluate the marginal likelihood on an adaptively selected
active set of predictors determined by the current shrinkage state. For each predictor
\(j\), we form the summary quantity
\[
v_j = \frac{1}{3}\left(\nu^{(\mathcal{A})}_j + \nu^{(\bar{\mathcal{A})}}_j + \nu^{(\delta)}_j\right),
\]
which averages the local variance components associated with that predictor across the
informative-source, noninformative-source, and contrast parameters. Predictors for which
\(v_j\) is very small are those simultaneously shrunk across all model components and thus
have negligible influence on the Gaussian integral. We therefore restrict computation to the
index set
\[
\mathcal{S} = \{\, j : v_j > \kappa \,\},
\]
for a fixed small threshold \(\kappa>0\) (say \(\kappa=0.1\)), and
evaluate the closed-form marginal likelihood using the design matrices restricted
to columns in \(\mathcal{S}\). Because coefficients outside \(\mathcal{S}\) are already
concentrated near zero under the prior, integrating over the reduced parameter space yields
a numerically equivalent contribution while substantially lowering computational cost.

To avoid permanently excluding predictors due to transient shrinkage early in the Markov
chain, a small subset of predictors are additionally retained at random at each evaluation.
This randomized retention step ensures that all coordinates are periodically reconsidered.

If \(s_{\text{eff}} = |\mathcal{S}|\) denotes the effective number of active predictors, the dominant
matrix factorizations scale as \(\mathcal{O}(s_{\text{eff}}^3)\) rather than \(\mathcal{O}(p^3)\), with
typically \(s_{\text{eff}} \ll p\). In practice, \(s_{\text{eff}}\) tracks the number of coefficients not strongly
shrunk by the global--local prior, so this strategy leverages sparsity already implied by
the model. This active-set restriction is used solely to accelerate marginal likelihood
evaluation within Algorithm 2.

\subsubsection{Tempering for the source study selection step}
\label{app:study_selection}

A key component of Algorithm~2 is the \emph{source study selection step}, 
in which the inclusion vector $\bm\gamma$ is updated to determine the partition 
of source studies into informative ($\mathcal A_{\bm\gamma}$) and 
noninformative ($\bar{\mathcal A}_{\bm\gamma}$) sets. 
At each iteration, for each source study $k = 1,\dots,K$, we evaluate the 
(unnormalized) log posterior mass of the current configuration 
$\bm\gamma$ and of a proposed flip $\bm\gamma'$,
\[
\log p_{\text{curr},k} = \log p(\bm\gamma \mid \mathcal D),
\qquad
\log p_{\text{prop},k} = \log p(\bm\gamma' \mid \mathcal D),
\]
where $\mathcal D$ denotes the collection of target and source data.

To improve mixing for $\gamma$ in high-dimensional settings, 
we employ \emph{tempering} during burn-in. Specifically, at iteration 
$i$ we replace the posterior by a tempered distribution
\[
p_{\kappa_i}(\bm\gamma)
\;\propto\;
p(\bm\gamma)^{\kappa_i},
\qquad \kappa_i \in (0,1],
\]
which is equivalent to rescaling the log posterior differences by $\kappa_i$ 
when computing Metropolis--Hastings acceptance probabilities. 
Tempering flattens the posterior landscape, reducing sharp likelihood contrasts 
between configurations and prevents stickiness or near-deterministic 
updates of $\bm\gamma$.

In practice, we apply tempering only during burn-in using the schedule
\[
\kappa_i = \frac{\sqrt{i+1}}{p},
\]
which gradually increases toward one, ensuring that the sampler ultimately 
targets the true posterior distribution. This strategy substantially improves 
mixing when the number of predictors $p$ is 
large.

\subsubsection{Degenerate source configurations}

Care must be taken for degenerate configurations  of the inclusion vector $\boldsymbol{\gamma}$ that may arise during MCMC sampling.  In particular, when the non-informative set is empty (i.e.,  $\boldsymbol{\gamma} = \mathbf{1}_K$), the subset  $\mathcal{D}_{\bar{\mathcal{A}}_{\boldsymbol{\gamma}}}$ contains no observations.  In this case, the likelihood contribution associated with  $\boldsymbol{w}^{(\bar{\mathcal{A}})}$ vanishes, and the corresponding full  conditional distributions are no longer informed by data.

A naive remedy is to temporarily impute  $\mathbf{X}^{(\bar{\mathcal{A}}_{\boldsymbol{\gamma}})}$ and $\mathbf{y}^{(\bar{\mathcal{A}}_{\boldsymbol{\gamma}})}$ with zeros.  While such zero-imputation is harmless for evaluating marginal likelihoods (since  empty subsets contribute neutrally) it is inappropriate within the Gibbs updates. Introducing artificial data in this manner effectively alters the likelihood and can induce spurious variability in the posterior samples, leading to unstable MCMC behavior. Instead, when $\mathcal{D}_{\bar{\mathcal{A}}_{\boldsymbol{\gamma}}}$ is empty, the regression parameter $\boldsymbol{w}^{(\bar{\mathcal{A}})}$ is set identically to $\bm{0}_{p}$ or sampled directly from its prior distribution.
Associated variance and shrinkage parameters may likewise be drawn from their prior distributions or fixed to particular values (see \ref{app:trans-dimensional} for an example). 

The other degenerate case occurs when the informative set is empty ($\boldsymbol{\gamma} = \mathbf{0}_K$) during sampling. In this setting, no modification to the sampling scheme is required, as inference reduces to a target-only analysis and the Gibbs updates for $\boldsymbol{w}^{(\mathcal{A})}$ and $\boldsymbol{\delta}$ remain well-defined.

%\paragraph*{(2) Mixing of the Latent Source Indicators} \\
%When the number of source datasets \(K\) is large, the sampler must explore acombinatorial model space of size \(2^{K}\), which can lead to slow mixing of the inclusion vector \(\boldsymbol{\gamma}\). This difficulty is amplified in high-dimensional settings, where large \(p\) may produce highly separated log marginal likelihood values across competing configurations, yielding near-deterministic Metropolis--Hastings updates and limited movement between models. To improve exploration during the early stages of sampling, we employ a tempered version of the posterior during burn-in that rescales the log-marginal likelihood contributions, thereby flattening the model space and facilitating more stable transitions among candidate informative sets. The target posterior is recovered after burn-in  (see Appendix~A for details).

\section{Horseshoe Prior BLAST Implementation Details}\label{app:hp_supp}

\subsection{Exact Blocked Metropolis-within-Gibbs Sampler for the Horseshoe Prior} \label{app:horseshoe_gibbs_sampler}

We describe the sampling steps for the exact blocked Metropolis-within-Gibbs algorithm from \citet{Johndrow2020ScalablePrior} targeting the posterior under the horseshoe prior. This scheme updates the local shrinkage parameters, the global shrinkage parameter, the error variance, and the regression coefficients.

We refer to the horseshoe prior specification from Section 2.3.3. Suppose we have an outcome vector $\mathbf{y}\in\mathbb{R}^{N\times1}$ and design matrix $\mathbf{X} \in \mathbb{R}^{N\times p}$. Let 
\begin{align*}
    \eta_j &= 1/\lambda_j^2\\
    \xi &= 1/ \tau^2 \\
    \mathbf{\Lambda} &= \operatorname{diag}(\eta_j^{-1}), \\
    \mathbf{M}_{\xi} &= I_N + \xi^{-1} \mathbf{X} \mathbf{\Lambda} \mathbf{X}^\top,
\end{align*}
and the marginal density of the global parameter \(\xi\) given \(\eta\) and $\psi$ is
\begin{align*}
    p(\xi \mid \eta, \psi) &= |\mathbf{M}_{\xi}|^{-1/2} \left( \frac{\omega}{2} + \frac{1}{2} \mathbf{\mathbf{y}}^\top \mathbf{M}_{\xi}^{-1} \mathbf{\mathbf{y}} \right)^{-(N + \omega)/2} \cdot \frac{1}{\sqrt{\frac{\xi}{\psi^2}}(1 + \xi  \psi^2)}.
\end{align*}
Each iteration of the sampler proceeds as follows:

\begin{align*}
    &\text{1. Sample } \eta_j \sim p(\eta_j \mid \xi, \bm\beta_j, \sigma^2) \propto \frac{1}{1 + \eta_j} \exp\left( -\frac{\bm\beta_j^2 \xi \eta_j}{2 \sigma^2} \right) \quad \text{for } j = 1, \ldots, p. \\
    &\text{2. Propose } \log(\xi^*) \sim \mathcal{N}(\log(\xi), s_{\xi}), \quad \text{accept with probability} \\
    &\quad \min\left(1, \frac{p(\xi^* \mid \eta)}{p(\xi \mid \eta)} \right). \\
    &\text{3. Sample } \sigma^2 \sim \text{InvGamma}\left( \frac{\omega + N}{2}, \frac{\omega + \mathbf{\mathbf{y}}^\top \mathbf{M}_{\xi}^{-1} \mathbf{\mathbf{y}}}{2} \right). \\
    &\text{4. Sample } \bm\beta \sim \mathcal{N}\left( (\mathbf{X}^\top \mathbf{X} + (\xi^{-1} \mathbf{\Lambda})^{-1})^{-1} \mathbf{X}^\top \mathbf{\mathbf{y}}, \, \sigma^2 (\mathbf{X}^\top \mathbf{X} + (\xi^{-1} \mathbf{\Lambda})^{-1})^{-1} \right).
\end{align*}

One may fix the scale parameter $\psi = 1$ or use the empirical Bayes strategies described in \ref{app:specifyglobalshrink} to obtain an estimate $\hat\psi$ for which subsequent Gibbs samples are produced. 

We denote a single Gibbs iteration using this sampler by drawing sequentially from each of the full conditionals above. The structure of \(\mathbf{M}_{\xi}\) and \(\mathbf{\Lambda}\) ensures computational efficiency for high-dimensional settings.

We also adopt the fast Gaussian sampling method of \citet{Bhattacharya2016FastRegression} for drawing \(\beta \mid \eta, \xi, \sigma^2\) efficiently. This method reduces the computational complexity from \(\mathcal{O}(p^3)\) to \(\mathcal{O}(N^2p)\), which is critical in large-$p$ regimes.

Given that \(\mathbf{M}_{\xi} = I_N + \xi^{-1} \mathbf{X} \mathbf{\Lambda} \mathbf{X}^\top\), the following steps are used to sample from the conditional distribution \(\bm\beta \mid \eta, \xi, \sigma^2\):

\begin{align*}
    &\text{1. Sample } u \sim \mathcal{N}(0, \xi^{-1} \mathbf{\Lambda}), \quad f \sim \mathcal{N}(0, \mathbf{I}_N) \text{ independently}, \\
    &\text{2. Set } v = \mathbf{X} u + f, \quad v^* = \mathbf{M}_{\xi}^{-1} \left( \frac{\mathbf{y}}{\sigma} - v \right), \\
    &\text{3. Set } \bm\beta = \sigma \left( u + \xi^{-1} \mathbf{\Lambda} \mathbf{X}^\top v^* \right).
\end{align*}
This formulation avoids the direct inversion of large \(p \times p\) matrices and instead uses matrix-vector operations involving \(N \times p\) and \(N \times N\) matrices. This is particularly advantageous when used in combination with the blocked Gibbs updates of the shrinkage and variance parameters described previously.

\subsection{Encouraging Greater Sparsity in Contrast Parameters}
\label{app:sparsity_contrasts}

In our high-dimensional multi-source transfer learning setting, it is often reasonable to assume that the contrast vector $\bm\delta$ is more sparse than the source coefficients $\bm{w}^{\mathcal{A}}$, since the target task is expected to differ from the source tasks only through a limited subset of features. Let $\xi_{w}$ and $\xi_{\delta}$ denote the global shrinkage parameters governing $\bm{w}^{\mathcal{A}}$ and $\bm\delta$, respectively. Under this intuition, one may wish to encourage stronger global shrinkage on the contrast coefficients by enforcing
\[
\xi_{\delta} > \xi_{w}.
\]

A convenient way to incorporate this preference within the sampler described in \ref{app:horseshoe_gibbs_sampler} is through an optional modification of the proposal distribution used to update $\xi_{\delta}$. Specifically, conditional on the current value of $\xi_{w}$, we restrict proposals for $\xi_{\delta}$ to lie above $\xi_{w}$. Working on the log scale, this corresponds to proposing
\[
\log(\xi_{\delta}^*) \sim 
\mathcal{N}\!\bigl(\log(\xi_{\delta}),\, s_{\delta}\bigr)\,
\mathbb{I}_{[\log(\xi_{w}),\,\infty)},
\]
that is, a random-walk normal proposal truncated below at $\log(\xi_{w})$. Here, the truncation boundary is treated as fixed within the Metropolis--Hastings update for $\xi_{\delta}$, conditional on the current iterate of $\xi_{w}$.

Because the proposal distribution is no longer symmetric, the Metropolis--Hastings acceptance probability must be adjusted to account for the truncation. In particular, the acceptance probability for a proposed move from $\xi_{\delta}$ to $\xi_{\delta}^*$ is given by
\[
\alpha
=
\min\!\left\{
1,\; \frac{p(\xi^* \mid \eta)}{p(\xi \mid \eta)} 
\cdot
\frac{
1-\Phi\!\left(\dfrac{\log(\xi_{w})-\log(\xi_{\delta})}{\sqrt{s}}\right)
}{
1-\Phi\!\left(\dfrac{\log(\xi_{w})-\log(\xi_{\delta}^*)}{\sqrt{s}}\right)
}
\right\},
\]
where $\Phi(\cdot)$ denotes the standard normal cumulative distribution function.

The additional ratio involving $\Phi(\cdot)$ corresponds to the ratio of normalizing constants for the truncated normal proposal and ensures detailed balance with respect to the target posterior distribution. Importantly, this modification does not alter the underlying model; it only biases the proposal mechanism to favor configurations in which the contrast parameters are subject to stronger global shrinkage than the source coefficients.

\subsection{Specifying the Global Shrinkage Parameter via Empirical Bayes Gibbs Sampling}\label{app:specifyglobalshrink}

Specifying the global shrinkage parameter $\tau$ in the horseshoe prior is critical for effective regularization. Common forms for the prior on $\tau$ include $\tau \sim \text{C}^+(0,1)$ \citep{Carvalho2009HandlingHorseshoe} or $\tau \sim \text{C}^+(0,\sigma)$ \citep{Carvalho2010TheSignals} where $\sigma$ is the error variance. However, specifying the prior using these formulations may have negative implications. \citet{Piironen2017SparsityPriors} caution that such choices can result in overly diffuse priors, particularly when $\tau$ is weakly identified by the data, leading to insufficient regularization and excessive prior mass on implausibly large values of $\tau$. Hence, they recommend setting $\tau \sim \text{C}^+(0, \psi^2)$ and propose the following functional form for $\psi$:
\begin{align}\label{psi_funcl_form}
    \psi = \frac{p_0}{p-p_0}\frac{1}{\sqrt{n}}
\end{align}
where $p_0$ is an initial guess of the number of signals, $\sigma$ is the error variance, $n$ is the sample size, and $p$ is the dimension of the regression parameter vector.  They show that this specification leads to more plausible prior distributions for the effective number of signals. However, the effectiveness of this approach depends on prior knowledge of the number of nonzero parameters, which may be unavailable in practice. Therefore, we propose the option of using an empirical Bayes approach to learn $p_0$ from the data. 

Since $p_0$ is a discrete, non-negative quantity, it is challenging to estimate directly in an empirical Bayes framework. Hence, $p_0$ can be reparametrized as $p_0 = p \times \text{logit}(\psi_0)$ and obtain the convenient form: $$\psi = \exp(\psi_0)\frac{1}{\sqrt{n}}$$ thereby reducing the problem to estimating $\psi_0$ which is unconstrained on the real line. An empirical Bayes estimate $\hat\psi_0$ can be  obtained by maximizing the marginal likelihood of the data
\begin{align}\label{psimarg}
    \hat{\psi}_0 = \argmax_{\psi_0}{p(\psi_0 | \mathbf{y})}
\end{align}

The computation of the marginal likelihood $p(\psi_0 | \mathbf{y})$ in (\ref{psimarg}) is intractable and requires numerical approximation.  However, given that our algorithms use Gibbs sampling, an empirical Bayes Gibbs sampling procedure \citep{Casella2001EmpiricalSampling} can be readily implemented to obtain the maximum likelihood estimate $\hat\psi_0$ without direct computation of this marginal likelihood. The estimation reduces to maximizing the log of the conditional likelihood via Monte Carlo Expectation-Maximization (EM) which will solve: 
\begin{align*}
    \hat\psi_0 = \argmax_{\psi_0}{\frac{1}{B}}\sum_{j = 1}^B\log{p(\psi_0| \tau^{2(j)})}
    % -\log{\left(\psi_{\sigma^{2(j)}} + \frac{1}{\tau^{2(j)}\psi_{\sigma^{2(j)}}}\right)}
\end{align*}
where the values $\tau^{2(j)}$ are obtained from $B$ iterations of a horseshoe Gibbs sampler (after burn-in). After convergence, we specify the prior on the global shrinkage as $\tau  \sim \text{C}^+(0, \hat\psi^2)$ where $\hat\psi = \exp(\hat\psi_0) \frac{1}{\sqrt{n}}$ and obtain subsequent Gibbs samples.

The empirical Bayes Gibbs sampler can be seamlessly integrated into Algorithm 1 following $M$ additional MCMC iterations after burn-in. Since the total sample size of the source studies $n_{\mathcal{A}} = \sum_k n_k$ is fixed, estimates of the hyperparameter for the source coefficients $\hat\psi^{\bm{w}} = \exp(\hat\psi_0^{\bm{w}})\frac{1}{\sqrt{n_{\mathcal{A}}}}$ and contrasts $\hat\psi^{\bm\delta} = \exp(\hat\psi_0^{\mathbf{\delta}})\frac{1}{\sqrt{n_{0}}}$ can be  separately determined from corresponding samples of error variances and shrinkage parameters. However, in Algorithm 2, the true sample size of the informative studies, $n_{\mathcal{A}_{\bm\gamma}} = \sum_{k \in \mathcal{A}_{\bm\gamma}}n_k$ is unknown due to dependence on a stochastic $\bm\gamma$, making the implementation slightly more nuanced. In this case, we recommend learning the overall sparsity from the $n_0$ samples in the target data alone. This is done by obtaining an estimate, $\hat\psi = \exp(\hat\psi_0)\frac{1}{\sqrt{n_0}}$, to give a rough approximation of the general sparsity of the regression coefficients. This estimate can be directly used as the scale parameter in the Half-Cauchy priors for the global shrinkage of the source coefficients and contrasts. 

\subsection{Degenerate inclusion configurations and numerical stabilization}
\label{app:trans-dimensional}

Our formulation avoids trans-dimensional exploration by maintaining a fixed-dimensional parameter space throughout sampling. In particular, all regression 
coefficients remain defined for every iteration of the algorithm, regardless of 
the current inclusion configuration. 

When the noninformative subset is empty 
($\bar{\mathcal{A}} = \emptyset$), the corresponding likelihood contribution 
vanishes. Consequently, parameters associated with this subset are not 
identified by the data and their posterior distributions reduce to their priors. 
In this setting, sampling $\bm{w}^{(\bar{\mathcal A})}$ directly from its prior 
is therefore both computationally convenient and probabilistically exact. 
Equivalently, one may deterministically fix 
$\bm{w}^{(\bar{\mathcal A})}=\mathbf 0$.

To further stabilize the sampler in these degenerate regimes, we regularize the 
associated global shrinkage and variance parameters. Specifically, we set
\[
\tau_{(\bar{\mathcal A})}
\sim
C^+(0,\eta)\,\mathbb I(\bar{\mathcal A}\neq\emptyset)
+
\mathcal D\!\left(\tfrac{1}{p^2}\right)\,\mathbb I(\bar{\mathcal A}=\emptyset),
\]
where $\mathcal D(\phi)$ denotes a Dirac mass at $\phi$. This choice effectively 
shrinks inactive coefficients to zero when no data are present. As for the local shrinkage parameters, they can be set to a fixed value of 1. Similarly, the error variance prior is modified to reflect pure sampling 
variability when the subset is empty:
\[
\sigma^2_{(\bar{\mathcal A})}
\sim
\mathrm{IG}(\alpha,\beta)\,\mathbb I(\bar{\mathcal A}\neq\emptyset)
+
\mathrm{IG}(\sqrt N+1,\, s\sqrt N)\,\mathbb I(\bar{\mathcal A}=\emptyset),
\]
where $s=\mathrm{Var}(\mathbf{y})$ (or $s=1$ for standardized outcomes) and $N$ is the total sample size. 

\section{Details of Asymptotic Behavior of Bayes Factors for Source Selection}

This appendix provides technical details underlying the Bayes factor results
stated in Section~3. We first review general asymptotic properties of Bayes
factors under standard likelihood regularity conditions and then specialize
these results to the BLAST framework, where competing models are indexed by different
source membership configurations $\boldsymbol{\gamma}$.

\subsection{General Bayes Factor Asymptotics}
Assume standard regularity conditions for likelihood-based model comparison,
including interior maximum likelihood estimators (MLEs), twice continuously
differentiable log-likelihoods, nonsingular Fisher information matrices,
remainder terms negligible on $O(n^{-1/2})$ neighborhoods, and priors that are
continuous and strictly positive in neighborhoods of the relevant estimators.
These conditions hold for ordinary linear regression and common generalized
linear models under standard design assumptions.

For a regular model $M_k$ with parameter $\theta \in \mathbb{R}^{d_k}$, let
$\ell_{n,k}(\theta)$ denote the log-likelihood based on $n$ observations and
$\hat{\theta}_k$ the corresponding MLE. A second-order Taylor expansion of
$\ell_{n,k}$ about $\hat{\theta}_k$ yields
\[
\ell_{n,k}(\theta)
=
\ell_{n,k}(\hat{\theta}_k)
-
\frac{1}{2}
(\theta-\hat{\theta}_k)^\top
H_{n,k}
(\theta-\hat{\theta}_k)
+
R_{n,k}(\theta),
\]
where $H_{n,k}=-\nabla^2\ell_{n,k}(\hat{\theta}_k)$ is the observed information
matrix and $R_{n,k}(\theta)=o_p(1)$ uniformly for
$\|\theta-\hat{\theta}_k\|=O(n^{-1/2})$.
Applying Laplace’s method and using $H_{n,k}\approx n I_k$, where $I_k$ denotes
the per-observation Fisher information, gives
\begin{equation}\label{eq:logm_app}
\log m_k(y)
=
\ell_{n,k}(\hat{\theta}_k)
-
\frac{d_k}{2}\log n
+
C_k
+
o_p(1),
\end{equation}
with
\[
C_k
=
\log \pi_k(\hat{\theta}_k)
+
\frac{d_k}{2}\log(2\pi)
-
\frac{1}{2}\log |I_k|,
\]
which is $O_p(1)$.

Comparing two models $M_0$ and $M_1$ yields the Bayes factor
\[
\mathrm{BF}_{10}
=
\frac{m_1(y)}{m_0(y)},
\]
and subtracting \eqref{eq:logm_app} gives
\begin{equation}\label{eq:logBF_app}
\log \mathrm{BF}_{10}
=
\Delta \ell_n
-
\frac{r}{2}\log n
+
O_p(1),
\end{equation}
where
\[
\Delta \ell_n
=
\ell_{n,1}(\hat{\theta}_1)
-
\ell_{n,0}(\hat{\theta}_0),
\qquad
r=d_1-d_0.
\]
Thus, asymptotically, the log Bayes factor decomposes into a likelihood fit term
$\Delta \ell_n$ and a complexity penalty $(r/2)\log n$, up to $O_p(1)$ constants
arising from priors and information determinants.

When $M_0$ is correctly specified and nested within $M_1$, Wilks’ theorem implies
$2\Delta \ell_n \xrightarrow{d} \chi^2_r$, so $\Delta \ell_n=O_p(1)$ and
$\log \mathrm{BF}_{10}\to -\infty$, yielding polynomial decay
$\mathrm{BF}_{10}=O_p(n^{-r/2})$ in favor of the smaller model.
Conversely, when the true data-generating distribution lies in $M_1$ but not in
$M_0$, by the law of large numbers and likelihood theory,the likelihood difference satisfies

$$
\frac{1}{n}\Delta\ell_n \xrightarrow{p} c := \mathbb{E}_{P_0}\big[\log p_1(Y\mid\theta_1^*)-\log p_0(Y\mid\theta_0^*)\big] > 0.
$$
 and the Bayes factor grows exponentially in $n$, favoring the
larger model.

\subsection{Bayes Factors for Source Selection in BLAST}
We now specialize these general results to the BLAST framework. Let
$\mathcal{S}=\{1,2,\ldots,K\}$ index the auxiliary studies, with study $k$ having
sample size $\eta_k$, and define the total auxiliary sample size
$n=\sum_{k=1}^K \eta_k$. A source membership configuration
$\boldsymbol{\gamma}\in\{0,1\}^K$ partitions $\mathcal{S}$ into an informative set
and a non-informative set, inducing a corresponding BLAST model with marginal
likelihood $m_{\boldsymbol{\gamma}}(\mathbf{y})$.

Consider two configurations $\boldsymbol{\gamma}^{(1)}$ and
$\boldsymbol{\gamma}^{(2)}$ and define the Bayes factor
\[
\mathrm{BF}_{12}
=
\frac{m_{\boldsymbol{\gamma}^{(1)}}(\mathbf{y})}
     {m_{\boldsymbol{\gamma}^{(2)}}(\mathbf{y})}.
\]
Applying the Laplace approximation as above yields
\begin{equation}\label{eq:logBF_gamma}
\log \mathrm{BF}_{12}
=
\Delta \ell_n
-
\frac{r}{2}\log n
+
O_p(1),
\end{equation}
where
$\Delta \ell_n
=
\ell_{n,1}(\hat{\theta}_1)-\ell_{n,2}(\hat{\theta}_2)$
and $r=d_1-d_2$.

Two distinct asymptotic cases arise, depending on whether the competing
configurations are non-nested or nested.

\paragraph{Non-nested configurations ($r=0$).}
If both configurations assign at least one auxiliary study to the informative
set and at least one to the non-informative set, the corresponding BLAST models
have equal parameter dimension, so $r=0$ and no complexity penalty appears.
Let $\ell_{\eta_k}$ denote the log-likelihood contribution of study $k$, and let
$\hat{\theta}_{\mathcal{A}}$ and $\hat{\theta}_{\bar{\mathcal{A}}}$ denote the MLEs
under informative and non-informative modeling, respectively. In this case,
\[
\Delta \ell_n
=
\sum_{k=1}^K
\bigl(
\gamma_k^{(1)}-\gamma_k^{(2)}
\bigr)
\Bigl\{
\ell_{\eta_k}(\hat{\theta}_{\mathcal{A}})
-
\ell_{\eta_k}(\hat{\theta}_{\bar{\mathcal{A}}})
\Bigr\},
\]
where $\gamma_k^{(1)}-\gamma_k^{(2)}\in\{-1,0,1\}$ indicates whether study $k$ is
treated differently across the two configurations.

Let $\theta_{\mathcal{A}}^*$ and $\theta_{\bar{\mathcal{A}}}^*$ denote the KL
projections (or true parameters under correct specification) associated with the
informative and non-informative parameterizations. By consistency of the MLE and
the law of large numbers,
\[
\frac{1}{n}\Delta \ell_n
\xrightarrow{p}
c
=
\sum_{k=1}^K
\bigl(
\gamma_k^{(1)}-\gamma_k^{(2)}
\bigr)
\,
\mathbb{E}_{P_0}
\!\left[
\log p(Y\mid\theta_{\mathcal{A}}^*)
-
\log p(Y\mid\theta_{\bar{\mathcal{A}}}^*)
\right].
\]
Consequently, $\Delta \ell_n = n c + o_p(n)$ and
$\log \mathrm{BF}_{12}=n c + O_p(1)$. If
$\boldsymbol{\gamma}^{(1)}$ is the true configuration, then $c>0$ and the Bayes
factor grows exponentially in $n$. By symmetry, if
$\boldsymbol{\gamma}^{(2)}$ is true, $\log \mathrm{BF}_{12}\to -\infty$.

\paragraph{Nested (boundary) configurations ($r\neq 0$).}
If one configuration assigns all auxiliary studies to either the informative or
the non-informative set, the resulting BLAST models are nested and differ in
parameter dimension. In this boundary case, the Bayes factor includes a
complexity penalty of order $(r/2)\log n$. When the smaller model is correctly
specified, Wilks’ theorem implies $\Delta \ell_n=O_p(1)$ and the Bayes factor
decays polynomially in $n$, favoring the simpler configuration. When the larger
model is correctly specified, $\Delta \ell_n=n c+o_p(n)$ for some $c>0$, and the
Bayes factor grows exponentially, with the fit advantage dominating the
complexity penalty.

\newpage
\bibliography{references.bib}